\documentclass{eptcs}
\providecommand{\event}{MSFP 2012}

\makeatletter
\@ifundefined{lhs2tex.lhs2tex.sty.read}%
  {\@namedef{lhs2tex.lhs2tex.sty.read}{}%
   \newcommand\SkipToFmtEnd{}%
   \newcommand\EndFmtInput{}%
   \long\def\SkipToFmtEnd#1\EndFmtInput{}%
  }\SkipToFmtEnd

\newcommand\ReadOnlyOnce[1]{\@ifundefined{#1}{\@namedef{#1}{}}\SkipToFmtEnd}
\usepackage{amstext}
\usepackage{amssymb}
\usepackage{stmaryrd}
\DeclareFontFamily{OT1}{cmtex}{}
\DeclareFontShape{OT1}{cmtex}{m}{n}
  {<5><6><7><8>cmtex8
   <9>cmtex9
   <10><10.95><12><14.4><17.28><20.74><24.88>cmtex10}{}
\DeclareFontShape{OT1}{cmtex}{m}{it}
  {<-> ssub * cmtt/m/it}{}

\DeclareFontShape{OT1}{cmtt}{bx}{n}
  {<5><6><7><8>cmtt8
   <9>cmbtt9
   <10><10.95><12><14.4><17.28><20.74><24.88>cmbtt10}{}
\DeclareFontShape{OT1}{cmtex}{bx}{n}
  {<-> ssub * cmtt/bx/n}{}

\newcommand{\Conid}[1]{\mathit{#1}}
\newcommand{\Varid}[1]{\mathit{#1}}
\newcommand{\anonymous}{\kern0.06em \vbox{\hrule\@width.5em}}

\usepackage{polytable}

\@ifundefined{mathindent}%
  {\newdimen\mathindent\mathindent\leftmargini}%
  {}%

\def\resethooks{%
  \global\let\SaveRestoreHook\empty
  \global\let\ColumnHook\empty}
\newcommand*{\savecolumns}[1][default]%
  {\g@addto@macro\SaveRestoreHook{\savecolumns[#1]}}
\newcommand*{\restorecolumns}[1][default]%
  {\g@addto@macro\SaveRestoreHook{\restorecolumns[#1]}}
\newcommand*{\aligncolumn}[2]%
  {\g@addto@macro\ColumnHook{\column{#1}{#2}}}

\resethooks

\newcommand{\onelinecommentchars}{\quad-{}- }
\newcommand{\commentbeginchars}{\enskip\{-}
\newcommand{\commentendchars}{-\}\enskip}

\newcommand{\visiblecomments}{%
  \let\onelinecomment=\onelinecommentchars
  \let\commentbegin=\commentbeginchars
  \let\commentend=\commentendchars}

\newcommand{\invisiblecomments}{%
  \let\onelinecomment=\empty
  \let\commentbegin=\empty
  \let\commentend=\empty}

\visiblecomments

\newlength{\blanklineskip}
\newcommand{\hsindent}[1]{\quad}
\let\hspre\empty
\let\hspost\empty

\EndFmtInput
\makeatother
\ReadOnlyOnce{polycode.fmt}%
\makeatletter

\newcommand{\hsnewpar}[1]%
  {{\parskip=0pt\parindent=0pt\par\vskip #1\noindent}}

\newcommand{\hscodestyle}{}

\newcommand{\sethscode}[1]%
  {\expandafter\let\expandafter\hscode\csname #1\endcsname
   \expandafter\let\expandafter\endhscode\csname end#1\endcsname}

  {\par\noindent
   \advance\leftskip\mathindent
   \hscodestyle
   \let\\=\@normalcr
   \let\hspre\(\let\hspost\)%
   \pboxed}%
  {\endpboxed\)%
   \par\noindent
   \ignorespacesafterend}

  {\hsnewpar\abovedisplayskip
   \advance\leftskip\mathindent
   \hscodestyle
   \let\hspre\(\let\hspost\)%
   \pboxed}%
  {\endpboxed%
   \hsnewpar\belowdisplayskip
   \ignorespacesafterend}

  {\hsnewpar\abovedisplayskip
   \advance\leftskip\mathindent
   \hscodestyle
   \let\\=\@normalcr
   \(\pboxed}%
  {\endpboxed\)%
   \hsnewpar\belowdisplayskip
   \ignorespacesafterend}

\newcommand{\plainhs}{\sethscode{plainhscode}}

\plainhs

  {\hsnewpar\abovedisplayskip
   \advance\leftskip\mathindent
   \hscodestyle
   \let\\=\@normalcr
   \(\parray}%
  {\endparray\)%
   \hsnewpar\belowdisplayskip
   \ignorespacesafterend}

  {\parray}{\endparray}

  {\(\parray}{\endparray\)}

\def\codeframewidth{\arrayrulewidth}
\RequirePackage{calc}

  {\parskip=\abovedisplayskip\par\noindent
   \hscodestyle
   \arrayrulewidth=\codeframewidth
   \tabular{@{}|p{\linewidth-2\arraycolsep-2\arrayrulewidth-2pt}|@{}}%
   \hline\framedhslinecorrect\\{-1.5ex}%
   \let\endoflinesave=\\
   \let\\=\@normalcr
   \(\pboxed}%
  {\endpboxed\)%
   \framedhslinecorrect\endoflinesave{.5ex}\hline
   \endtabular
   \parskip=\belowdisplayskip\par\noindent
   \ignorespacesafterend}

\newcommand{\framedhslinecorrect}[2]%
  {#1[#2]}

  {\(\def\column##1##2{}%
   \let\>\undefined\let\<\undefined\let\\\undefined
   \newcommand\>[1][]{}\newcommand\<[1][]{}\newcommand\\[1][]{}%
   \def\fromto##1##2##3{##3}%
   }{\) }%

  {\let\orighscode=\hscode
   \let\origendhscode=\endhscode
   \def\endhscode{\def\hscode{\endgroup\def\@currenvir{hscode}\\}\begingroup}
   \orighscode\def\hscode{\endgroup\def\@currenvir{hscode}}}%
  {\origendhscode
   \global\let\hscode=\orighscode
   \global\let\endhscode=\origendhscode}%

\makeatother
\EndFmtInput
\ReadOnlyOnce{agda.fmt}%

\RequirePackage[T1]{fontenc}
\RequirePackage[utf8x]{inputenc}
\RequirePackage{ucs}
\RequirePackage{amsfonts}

\DeclareUnicodeCharacter{737}{\textsuperscript{l}}
\DeclareUnicodeCharacter{8718}{\ensuremath{\blacksquare}}
\DeclareUnicodeCharacter{8759}{::}
\DeclareUnicodeCharacter{9669}{\ensuremath{\triangleleft}}
\DeclareUnicodeCharacter{8799}{\ensuremath{\stackrel{\scriptscriptstyle ?}{=}}}
\DeclareUnicodeCharacter{10214}{\ensuremath{\llbracket}}
\DeclareUnicodeCharacter{10215}{\ensuremath{\rrbracket}}

\renewcommand\Varid[1]{\mathord{\textsf{#1}}}
\let\Conid\Varid
\newcommand\Keyword[1]{\textsf{\textbf{#1}}}
\EndFmtInput

\usepackage[numbers]{natbib}

\newcommand{\hidden}[1]{\!\!}

\title{From Mathematics to Abstract Machine \\[0.5ex] \Large{A formal derivation of an executable Krivine machine}}
\author{Wouter Swierstra
\institute{Radboud University Nijmegen}\\
\email{w.swierstra@cs.ru.nl}}

\def\titlerunning{From Mathematics to Abstract Machine}
\def\authorrunning{W.~Swierstra}

\begin{document}
\maketitle

\begin{abstract}
  This paper presents the derivation of an executable Krivine abstract
  machine from a small step interpreter for the simply typed lambda
  calculus in the dependently typed programming language Agda.
\end{abstract}

\section{Introduction}
\label{sec:introduction}

There is a close relationship between lambda calculi with explicit
substitutions and abstract machines. Biernacka and
Danvy~\cite{biernacka} have shown how to derive several well-known
abstract machines including the Krivine
machine~\citep{cregut,curien,hankin,krivine}, the CEK
machine~\citep{cek}, and the Zinc machine~\citep{zinc}. Starting with
a functional program that evaluates by repeated head reduction, each
of these abstract machines may be derived by a series of program
transformations. Every transformation is carefully motivated in the
accompanying text. This paper aims to nail down the correctness of
these derivations further and, in the process, uncover even more
structure.

In this paper we show how the derivation presented by Biernacka and
Danvy can be formalized in the dependently typed programming language
Agda~\citep{norell}. What do we hope to gain by doing so?  In their
study relating evaluators and abstract machines, Ager et
al.~\cite{ager} state in the introduction:
\begin{quote}
  Most of our implementations of the abstract machines raise compiler
  warnings about non-exhaustive matches. These are inherent to
  programming abstract machines in an ML-like language.
\end{quote}
This paper demonstrates that these non-exhaustive matches are
\emph{not} inherent to a dependently typed programming language such
as Agda. All the functions we present here are structurally recursive
and provide alternatives for every case branch. This shift to a
dependently typed language gives us many properties of evaluation `for
free.' For example, from the types alone we learn that evaluation is
type preserving and that every term can be decomposed uniquely into a
redex and evaluation context. Finally, using Agda enables us to
provide a \emph{machine-checked proof} of the correctness of every
transformation. More specifically, this paper makes the following
concrete contributions:

\begin{itemize}
\item We describe the implementation of a small step evaluator in Agda
  that normalizes by repeated head reduction
  (Section~\ref{sec:reduction}). To convince Agda's termination
  checker that our definition is sound, we provide a normalization
  proof in the style of Tait~\cite{tait}, originally sketched by
  Coquand~\cite{coquand} (Section~\ref{sec:iterated-head-reduction}).
\item Applying the \emph{refocusing} transformation~\citep{refocus},
  yields a small-step abstract machine that is not yet tail-recursive
  (Section~\ref{sec:refocusing}). We prove that this transformation
  preserves the semantics and termination properties of the small-step
  evaluator from Section~\ref{sec:iterated-head-reduction}.
\item This small-step abstract machine can be transformed further to derive
  the Krivine machine (Section~\ref{sec:krivine}). Once again, we show
  that the transformation preserves the semantics and termination
  properties of the small-step abstract machine from
  Section~\ref{sec:refocusing}.
\end{itemize}

This paper is a literate Agda program. Rather than spelling out the
details of every proof, we will only sketch the necessary lemmas and
definitions. The complete source code, including proofs, is available
online.\footnote{The source code, compatible with Agda version 2.3,
  is available from \url{http://www.cs.ru.nl/~wouters}.} Every section
in this paper defines a separate module, allowing us to reuse the same
names for the functions and data types presented in individual
sections. Finally, the code in this paper uses a short Agda Prelude
that is included in an appendix. Readers unfamiliar with Agda may want
to consult one of the many tutorials and introductions that are
available~\cite{peter-ana:tutorial,norell:tutorial,oury-swierstra:power-of-pi}.

\section{Types and terms}
\label{sec:types-and-terms}

Before we can develop the series of evaluators, we need to define the
terms and types of the simply typed lambda calculus.

\begin{hscode}\SaveRestoreHook
\column{B}{@{}>{\hspre}l<{\hspost}@{}}%
\column{3}{@{}>{\hspre}l<{\hspost}@{}}%
\column{5}{@{}>{\hspre}l<{\hspost}@{}}%
\column{E}{@{}>{\hspre}l<{\hspost}@{}}%
\>[3]{}\Keyword{data}\;\Conid{Ty}\;\mathbin{:}\;\Conid{Set}\;\Keyword{where}{}\<[E]%
\\
\>[3]{}\hsindent{2}{}\<[5]%
\>[5]{}\Conid{O}\;\mathbin{:}\;\Conid{Ty}{}\<[E]%
\\
\>[3]{}\hsindent{2}{}\<[5]%
\>[5]{}\_\!\Rightarrow\!\_\;\mathbin{:}\;\Conid{Ty}\;\to \;\Conid{Ty}\;\to \;\Conid{Ty}{}\<[E]%
\\[\blanklineskip]%
\>[3]{}\Conid{Context}\;\mathbin{:}\;\Conid{Set}{}\<[E]%
\\
\>[3]{}\Conid{Context}\;\mathrel{=}\;\Conid{List}\;\Conid{Ty}{}\<[E]%
\ColumnHook
\end{hscode}\resethooks
The data type \ensuremath{\Conid{Ty}} represents the types of the simply typed lambda
calculus with one base type \ensuremath{\Conid{O}}. A context is defined to be a list of
types. Typically the variables \ensuremath{\sigma } and \ensuremath{\tau } range over
types; the variables \ensuremath{\Gamma } and \ensuremath{\Delta } range over contexts.

Next we define the data types of well-typed, well-scoped variables and
lambda terms:
\begin{hscode}\SaveRestoreHook
\column{B}{@{}>{\hspre}l<{\hspost}@{}}%
\column{3}{@{}>{\hspre}l<{\hspost}@{}}%
\column{5}{@{}>{\hspre}l<{\hspost}@{}}%
\column{E}{@{}>{\hspre}l<{\hspost}@{}}%
\>[3]{}\Keyword{data}\;\Conid{Ref}\;\mathbin{:}\;\Conid{Context}\;\to \;\Conid{Ty}\;\to \;\Conid{Set}\;\Keyword{where}{}\<[E]%
\\
\>[3]{}\hsindent{2}{}\<[5]%
\>[5]{}\Conid{Top}\;\mathbin{:}\;\hidden{\Gamma \;\sigma }\;\Conid{Ref}\;(\Conid{Cons}\;\sigma \;\Gamma )\;\sigma {}\<[E]%
\\
\>[3]{}\hsindent{2}{}\<[5]%
\>[5]{}\Conid{Pop}\;\mathbin{:}\;\hidden{\Gamma \;\sigma \;\tau }\;\Conid{Ref}\;\Gamma \;\sigma \;\to \;\Conid{Ref}\;(\Conid{Cons}\;\tau \;\Gamma )\;\sigma {}\<[E]%
\\[\blanklineskip]%
\>[3]{}\Keyword{data}\;\Conid{Term}\;\mathbin{:}\;\Conid{Context}\;\to \;\Conid{Ty}\;\to \;\Conid{Set}\;\Keyword{where}{}\<[E]%
\\
\>[3]{}\hsindent{2}{}\<[5]%
\>[5]{}\Conid{Lam}\;\mathbin{:}\;\hidden{\Gamma \;\sigma \;\tau }\;\Conid{Term}\;(\Conid{Cons}\;\sigma \;\Gamma )\;\tau \;\to \;\Conid{Term}\;\Gamma \;(\sigma \;\Rightarrow\;\tau ){}\<[E]%
\\
\>[3]{}\hsindent{2}{}\<[5]%
\>[5]{}\Conid{App}\;\mathbin{:}\;\hidden{\Gamma \;\sigma \;\tau }\;\Conid{Term}\;\Gamma \;(\sigma \;\Rightarrow\;\tau )\;\to \;\Conid{Term}\;\Gamma \;\sigma \;\to \;\Conid{Term}\;\Gamma \;\tau {}\<[E]%
\\
\>[3]{}\hsindent{2}{}\<[5]%
\>[5]{}\Conid{Var}\;\mathbin{:}\;\hidden{\Gamma \;\sigma }\;\Conid{Ref}\;\Gamma \;\sigma \;\to \;\Conid{Term}\;\Gamma \;\sigma {}\<[E]%
\ColumnHook
\end{hscode}\resethooks
These definitions are entirely standard.
There are three constructors for the simply typed lambda calculus:
\ensuremath{\Conid{Lam}} introduces a lambda, extending the context; the \ensuremath{\Conid{App}}
constructor applies a term of type \ensuremath{\sigma \;\Rightarrow\;\tau } to an argument of type
\ensuremath{\sigma }; the \ensuremath{\Conid{Var}} constructor references a variable bound in the context.

Note that in the typeset code presented in this paper, any unbound
variables in type signatures are implicitly universally quantified, as
is the convention in Haskell~\cite{view} and
Epigram~\cite{haskell98}. When we wish to be more explicit about
implicit arguments, we will adhere to Agda's notation of enclosing
such arguments in curly braces.

Next, we can define the data types representing \emph{closed} terms. A
\emph{closure} is a term \ensuremath{\Varid{t}} paired with an environment containing
closed terms for all the free variables in \ensuremath{\Varid{t}}. Furthermore, closed
terms are closed under application. This yields the two
mutually recursive data types defined below.
\begin{hscode}\SaveRestoreHook
\column{B}{@{}>{\hspre}l<{\hspost}@{}}%
\column{5}{@{}>{\hspre}l<{\hspost}@{}}%
\column{7}{@{}>{\hspre}l<{\hspost}@{}}%
\column{E}{@{}>{\hspre}l<{\hspost}@{}}%
\>[5]{}\Keyword{data}\;\Conid{Closed}\;\mathbin{:}\;\Conid{Ty}\;\to \;\Conid{Set}\;\Keyword{where}{}\<[E]%
\\
\>[5]{}\hsindent{2}{}\<[7]%
\>[7]{}\Conid{Closure}\;\mathbin{:}\;\hidden{\Gamma \;\sigma }\;\Conid{Term}\;\Gamma \;\sigma \;\to \;\Conid{Env}\;\Gamma \;\to \;\Conid{Closed}\;\sigma {}\<[E]%
\\
\>[5]{}\hsindent{2}{}\<[7]%
\>[7]{}\Conid{Clapp}\;\mathbin{:}\;\hidden{\sigma \;\tau }\;\Conid{Closed}\;(\sigma \;\Rightarrow\;\tau )\;\to \;\Conid{Closed}\;\sigma \;\to \;\Conid{Closed}\;\tau {}\<[E]%
\\[\blanklineskip]%
\>[5]{}\Keyword{data}\;\Conid{Env}\;\mathbin{:}\;\Conid{Context}\;\to \;\Conid{Set}\;\Keyword{where}{}\<[E]%
\\
\>[5]{}\hsindent{2}{}\<[7]%
\>[7]{}\Conid{Nil}\;\mathbin{:}\;\Conid{Env}\;\Conid{Nil}{}\<[E]%
\\
\>[5]{}\hsindent{2}{}\<[7]%
\>[7]{}\Varid{\char95 ·\char95 }\;\mathbin{:}\;\hidden{\Gamma \;\sigma }\;\Conid{Closed}\;\sigma \;\to \;\Conid{Env}\;\Gamma \;\to \;\Conid{Env}\;(\Conid{Cons}\;\sigma \;\Gamma ){}\<[E]%
\ColumnHook
\end{hscode}\resethooks
This is a variation of Curien's $\lambda\rho$-calculus, proposed by
Biernacka and Danvy~\cite{biernacka}. A similar choice of closed terms
was independently proposed by Coquand~\cite{coquand}.

The aim of evaluation is to compute a \emph{value} for every closed
term. Closed lambda expressions are the only values in our
language. The final definitions in this section capture this:
\begin{hscode}\SaveRestoreHook
\column{B}{@{}>{\hspre}l<{\hspost}@{}}%
\column{3}{@{}>{\hspre}l<{\hspost}@{}}%
\column{5}{@{}>{\hspre}l<{\hspost}@{}}%
\column{E}{@{}>{\hspre}l<{\hspost}@{}}%
\>[3]{}\Varid{isVal}\;\mathbin{:}\;\hidden{\sigma }\;\Conid{Closed}\;\sigma \;\to \;\Conid{Set}{}\<[E]%
\\
\>[3]{}\Varid{isVal}\;(\Conid{Closure}\;(\Conid{Lam}\;\Varid{body})\;\Varid{env})\;\mathrel{=}\;\Conid{Unit}{}\<[E]%
\\
\>[3]{}\Varid{isVal}\;\anonymous \;\mathrel{=}\;\Conid{Empty}{}\<[E]%
\\[\blanklineskip]%
\>[3]{}\Keyword{data}\;\Conid{Value}\;(\sigma \;\mathbin{:}\;\Conid{Ty})\;\mathbin{:}\;\Conid{Set}\;\Keyword{where}{}\<[E]%
\\
\>[3]{}\hsindent{2}{}\<[5]%
\>[5]{}\Conid{Val}\;\mathbin{:}\;(\Varid{c}\;\mathbin{:}\;\Conid{Closed}\;\sigma )\;\to \;\Varid{isVal}\;\Varid{c}\;\to \;\Conid{Value}\;\sigma {}\<[E]%
\ColumnHook
\end{hscode}\resethooks

With these types in place, we can specify the type of the evaluation
function we will define in the coming sections:
\begin{hscode}\SaveRestoreHook
\column{B}{@{}>{\hspre}l<{\hspost}@{}}%
\column{3}{@{}>{\hspre}l<{\hspost}@{}}%
\column{E}{@{}>{\hspre}l<{\hspost}@{}}%
\>[3]{}\Varid{evaluate}\;\mathbin{:}\;\Conid{Closed}\;\sigma \;\to \;\Conid{Value}\;\sigma {}\<[E]%
\ColumnHook
\end{hscode}\resethooks
\section{Reduction}
\label{sec:reduction}

Writing $t \, [ \mathit{env} ]$ to denote the closure consisting of a
term $t$ and an environment $\mathit{env}$, the four rules in
below specify a normal-order small step
reduction relation for the closed terms. In this section, we will
start to implement these rules in Agda.

\begin{figure}[h]
  \centering

  \textsc{Lookup} \hspace{1em} $i \, [c_1,c_2, \ldots c_n] \to c_i$
  \vspace{2ex}

  \textsc{App} \hspace{1em} $(t_0 \; t_1) \, [ \mathit{env} ] \to (t_0 \, [ \mathit{env} ]) \; (t_1 \, [ \mathit{env} ])$
  \vspace{2ex}

  \textsc{Beta} \hspace{1em} $((\lambda t) \, [\mathit{env}]) \; x \to t \, [x \cdot \mathit{env}]$
  \vspace{2ex}

  \textsc{Left} \hspace{1em} if $c_0 \to c_0'$ then $c_0 \; c_1 \to c_0' \; c_1$
  \vspace{2ex}

  \label{fig:small-step}
\end{figure}

In the style of Danvy and Nielsen~\cite{refocus}, we define a single
reduction step in three parts. First, we decompose a closed term into
a redex and an evaluation context. Second, we contract the redex to
form a new closed term. Finally, we plug the resulting closed term
back into the evaluation context.

To define such a three-step reduction step, we start by defining the
\ensuremath{\Conid{Redex}} type, corresponding to the left-hand sides of the first three
rules above.

\begin{hscode}\SaveRestoreHook
\column{B}{@{}>{\hspre}l<{\hspost}@{}}%
\column{3}{@{}>{\hspre}l<{\hspost}@{}}%
\column{5}{@{}>{\hspre}l<{\hspost}@{}}%
\column{E}{@{}>{\hspre}l<{\hspost}@{}}%
\>[3]{}\Keyword{data}\;\Conid{Redex}\;\mathbin{:}\;\Conid{Ty}\;\to \;\Conid{Set}\;\Keyword{where}{}\<[E]%
\\
\>[3]{}\hsindent{2}{}\<[5]%
\>[5]{}\Conid{Lookup}\;\mathbin{:}\;\hidden{\Gamma \;\sigma }\;\Conid{Ref}\;\Gamma \;\sigma \;\to \;\Conid{Env}\;\Gamma \;\to \;\Conid{Redex}\;\sigma {}\<[E]%
\\
\>[3]{}\hsindent{2}{}\<[5]%
\>[5]{}\Conid{Rapp}\;\mathbin{:}\;\hidden{\Gamma \;\sigma \;\tau }\;\Conid{Term}\;\Gamma \;(\sigma \;\Rightarrow\;\tau )\;\to \;\Conid{Term}\;\Gamma \;\sigma \;\to \;\Conid{Env}\;\Gamma \;\to \;\Conid{Redex}\;\tau {}\<[E]%
\\
\>[3]{}\hsindent{2}{}\<[5]%
\>[5]{}\Conid{Beta}\;\mathbin{:}\;\hidden{\Gamma \;\sigma \;\tau }\;\Conid{Term}\;(\Conid{Cons}\;\sigma \;\Gamma )\;\tau \;\to \;\Conid{Env}\;\Gamma \;\to \;\Conid{Closed}\;\sigma \;\to \;\Conid{Redex}\;\tau {}\<[E]%
\ColumnHook
\end{hscode}\resethooks
\newpage
\noindent Of course, every redex can be mapped back to the closed term that it represents.

\begin{hscode}\SaveRestoreHook
\column{B}{@{}>{\hspre}l<{\hspost}@{}}%
\column{3}{@{}>{\hspre}l<{\hspost}@{}}%
\column{E}{@{}>{\hspre}l<{\hspost}@{}}%
\>[3]{}\Varid{fromRedex}\;\mathbin{:}\;\hidden{\sigma }\;\Conid{Redex}\;\sigma \;\to \;\Conid{Closed}\;\sigma {}\<[E]%
\\
\>[3]{}\Varid{fromRedex}\;(\Conid{Lookup}\;\Varid{i}\;\Varid{env})\;\mathrel{=}\;\Conid{Closure}\;(\Conid{Var}\;\Varid{i})\;\Varid{env}{}\<[E]%
\\
\>[3]{}\Varid{fromRedex}\;(\Conid{Rapp}\;\Varid{f}\;\Varid{x}\;\Varid{env})\;\mathrel{=}\;\Conid{Closure}\;(\Conid{App}\;\Varid{f}\;\Varid{x})\;\Varid{env}{}\<[E]%
\\
\>[3]{}\Varid{fromRedex}\;(\Conid{Beta}\;\Varid{body}\;\Varid{env}\;\Varid{arg})\;\mathrel{=}\;\Conid{Clapp}\;(\Conid{Closure}\;(\Conid{Lam}\;\Varid{body})\;\Varid{env})\;\Varid{arg}{}\<[E]%
\ColumnHook
\end{hscode}\resethooks

Next, we define the \ensuremath{\Varid{contract}} function that computes the result of
contracting a single redex:

\begin{hscode}\SaveRestoreHook
\column{B}{@{}>{\hspre}l<{\hspost}@{}}%
\column{3}{@{}>{\hspre}l<{\hspost}@{}}%
\column{12}{@{}>{\hspre}l<{\hspost}@{}}%
\column{E}{@{}>{\hspre}l<{\hspost}@{}}%
\>[3]{}\Varid{\char95 !\char95 }\;\mathbin{:}\;\hidden{\Gamma \;\sigma }\;\Conid{Env}\;\Gamma \;\to \;\Conid{Ref}\;\Gamma \;\sigma \;\to \;\Conid{Closed}\;\sigma {}\<[E]%
\\
\>[3]{}\Conid{Nil}\;\mathbin{!}\;(){}\<[E]%
\\
\>[3]{}(\Varid{x}\;\Varid{·}\;\anonymous )\;{}\<[12]%
\>[12]{}\mathbin{!}\;\Conid{Top}\;\mathrel{=}\;\Varid{x}{}\<[E]%
\\
\>[3]{}(\Varid{x}\;\Varid{·}\;\Varid{xs})\;\mathbin{!}\;\Conid{Pop}\;\Varid{r}\;\mathrel{=}\;\Varid{xs}\;\mathbin{!}\;\Varid{r}{}\<[E]%
\\[\blanklineskip]%
\>[3]{}\Varid{contract}\;\mathbin{:}\;\hidden{\sigma }\;\Conid{Redex}\;\sigma \;\to \;\Conid{Closed}\;\sigma {}\<[E]%
\\
\>[3]{}\Varid{contract}\;(\Conid{Lookup}\;\Varid{i}\;\Varid{env})\;\mathrel{=}\;\Varid{env}\;\mathbin{!}\;\Varid{i}{}\<[E]%
\\
\>[3]{}\Varid{contract}\;(\Conid{Rapp}\;\Varid{f}\;\Varid{x}\;\Varid{env})\;\mathrel{=}\;\Conid{Clapp}\;(\Conid{Closure}\;\Varid{f}\;\Varid{env})\;(\Conid{Closure}\;\Varid{x}\;\Varid{env}){}\<[E]%
\\
\>[3]{}\Varid{contract}\;(\Conid{Beta}\;\Varid{body}\;\Varid{env}\;\Varid{arg})\;\mathrel{=}\;\Conid{Closure}\;\Varid{body}\;(\Varid{arg}\;\Varid{·}\;\Varid{env}){}\<[E]%
\ColumnHook
\end{hscode}\resethooks
In the \ensuremath{\Conid{Lookup}} case, we look up the variable from the environment
using the \ensuremath{\Varid{\char95 !\char95 }} operator. The \ensuremath{\Conid{Rapp}} case distributes the environment
over the two terms. Finally, \ensuremath{\Conid{Beta}} reduction extends the environment
with the argument \ensuremath{\Varid{arg}}, and uses the extended environment to create a
new closure from the body of a lambda. Once again, the definition of
the \ensuremath{\Varid{contract}} function closely follows the first three reduction
rules that we formulated above.

While this describes how to contract a single redex, we still need to
define the \emph{decomposition} of a term into a redex and a reduction
context. We begin by defining an evaluation context as the list of
arguments encountered along the spine of a term:

\begin{hscode}\SaveRestoreHook
\column{B}{@{}>{\hspre}l<{\hspost}@{}}%
\column{3}{@{}>{\hspre}l<{\hspost}@{}}%
\column{5}{@{}>{\hspre}l<{\hspost}@{}}%
\column{E}{@{}>{\hspre}l<{\hspost}@{}}%
\>[3]{}\Keyword{data}\;\Conid{EvalContext}\;\mathbin{:}\;\Conid{Ty}\;\to \;\Conid{Ty}\;\to \;\Conid{Set}\;\Keyword{where}{}\<[E]%
\\
\>[3]{}\hsindent{2}{}\<[5]%
\>[5]{}\Conid{MT}\;\mathbin{:}\;\hidden{\sigma }\;\Conid{EvalContext}\;\sigma \;\sigma {}\<[E]%
\\
\>[3]{}\hsindent{2}{}\<[5]%
\>[5]{}\Conid{ARG}\;\mathbin{:}\;\hidden{\sigma \;\tau \;\rho }\;\Conid{Closed}\;\sigma \;\to \;\Conid{EvalContext}\;\tau \;\rho \;\to \;\Conid{EvalContext}\;(\sigma \;\Rightarrow\;\tau )\;\rho {}\<[E]%
\ColumnHook
\end{hscode}\resethooks

Ignoring the \ensuremath{\Conid{Ty}} indices for the moment, an evaluation context is
simply a list of closed terms. Given any evaluation context \ensuremath{\Varid{ctx}} and
term \ensuremath{\Varid{t}}, we would like to plug \ensuremath{\Varid{t}} in the context by iteratively
applying \ensuremath{\Varid{t}} to all the arguments in \ensuremath{\Varid{ctx}}. For this to type check,
the term \ensuremath{\Varid{t}} should abstract over all the variables in the evaluation
context. We enforce this by indexing the \ensuremath{\Conid{EvalContext}} type by the
`source' and `destination' types in the style of
Atkey~\cite{atkey}. The \ensuremath{\Varid{plug}} operation itself then applies any
arguments from the evaluation context to its argument term:

\begin{hscode}\SaveRestoreHook
\column{B}{@{}>{\hspre}l<{\hspost}@{}}%
\column{3}{@{}>{\hspre}l<{\hspost}@{}}%
\column{E}{@{}>{\hspre}l<{\hspost}@{}}%
\>[3]{}\Varid{plug}\;\mathbin{:}\;\hidden{\sigma \;\tau }\;\Conid{EvalContext}\;\sigma \;\tau \;\to \;\Conid{Closed}\;\sigma \;\to \;\Conid{Closed}\;\tau {}\<[E]%
\\
\>[3]{}\Varid{plug}\;\Conid{MT}\;\Varid{f}\;\mathrel{=}\;\Varid{f}{}\<[E]%
\\
\>[3]{}\Varid{plug}\;(\Conid{ARG}\;\Varid{x}\;\Varid{ctx})\;\Varid{f}\;\mathrel{=}\;\Varid{plug}\;\Varid{ctx}\;(\Conid{Clapp}\;\Varid{f}\;\Varid{x}){}\<[E]%
\ColumnHook
\end{hscode}\resethooks

Finally, we define the decomposition of a closed term into a redex and
evaluation context as a \emph{view}~\citep{view,wadler} on closed
terms. Defining such a view consists of two parts: a data type
\ensuremath{\Conid{Decomposition}} indexed by a closed term, and a function
\ensuremath{\Varid{decompose}} that maps every closed term to its \ensuremath{\Conid{Decomposition}}.

We will start by defining a data type \ensuremath{\Conid{Decomposition}}. There are two
constructors, corresponding to the two possible outcomes of
decomposing a closed term \ensuremath{\Varid{c}}: either \ensuremath{\Varid{c}} is a value, in which case we
have the closure of a \ensuremath{\Conid{Lam}}-term and an environment; alternatively,
\ensuremath{\Varid{c}} can be decomposed into a redex \ensuremath{\Varid{r}} and an evaluation context
\ensuremath{\Varid{ctx}}, such that plugging the term corresponding to \ensuremath{\Varid{r}} in the
evaluation context \ensuremath{\Varid{ctx}} is equal to the original term \ensuremath{\Varid{c}}:
\begin{hscode}\SaveRestoreHook
\column{B}{@{}>{\hspre}l<{\hspost}@{}}%
\column{3}{@{}>{\hspre}l<{\hspost}@{}}%
\column{6}{@{}>{\hspre}l<{\hspost}@{}}%
\column{8}{@{}>{\hspre}l<{\hspost}@{}}%
\column{E}{@{}>{\hspre}l<{\hspost}@{}}%
\>[3]{}\Keyword{data}\;\Conid{Decomposition}\;\mathbin{:}\;\hidden{\sigma }\;\Conid{Closed}\;\sigma \;\to \;\Conid{Set}\;\Keyword{where}{}\<[E]%
\\
\>[3]{}\hsindent{3}{}\<[6]%
\>[6]{}\Conid{Val}\;\mathbin{:}\;\hidden{\sigma \;\tau \;\Gamma }\;(\Varid{body}\;\mathbin{:}\;\Conid{Term}\;(\Conid{Cons}\;\sigma \;\Gamma )\;\tau )\;\to \;(\Varid{env}\;\mathbin{:}\;\Conid{Env}\;\Gamma )\;\to \;{}\<[E]%
\\
\>[6]{}\hsindent{2}{}\<[8]%
\>[8]{}\Conid{Decomposition}\;(\Conid{Closure}\;(\Conid{Lam}\;\Varid{body})\;\Varid{env}){}\<[E]%
\\
\>[3]{}\hsindent{3}{}\<[6]%
\>[6]{}\Conid{Decompose}\;\mathbin{:}\;\hidden{\tau \;\sigma }\;(\Varid{r}\;\mathbin{:}\;\Conid{Redex}\;\sigma )\;\to \;(\Varid{ctx}\;\mathbin{:}\;\Conid{EvalContext}\;\sigma \;\tau )\;\to \;{}\<[E]%
\\
\>[6]{}\hsindent{2}{}\<[8]%
\>[8]{}\Conid{Decomposition}\;(\Varid{plug}\;\Varid{ctx}\;(\Varid{fromRedex}\;\Varid{r})){}\<[E]%
\ColumnHook
\end{hscode}\resethooks

Next we show how every closed term \ensuremath{\Varid{c}} can be decomposed into a
\ensuremath{\Conid{Decomposition}\;\Varid{c}}. We do so by defining a pair of 
functions, \ensuremath{\Varid{load}} and \ensuremath{\Varid{unload}}. The \ensuremath{\Varid{load}} function traverses the
spine of \ensuremath{\Varid{c}}, accumulating any arguments we encounter in an evaluation
context until we find a redex or a closure containing a \ensuremath{\Conid{Lam}}. The
\ensuremath{\Varid{unload}} function inspects the evaluation context that \ensuremath{\Varid{load}} has
accumulated in order to decide if a lambda is indeed a value, or
whether it still has further arguments, and hence corresponds to a
\ensuremath{\Conid{Beta}} redex:

\begin{hscode}\SaveRestoreHook
\column{B}{@{}>{\hspre}l<{\hspost}@{}}%
\column{5}{@{}>{\hspre}l<{\hspost}@{}}%
\column{7}{@{}>{\hspre}l<{\hspost}@{}}%
\column{E}{@{}>{\hspre}l<{\hspost}@{}}%
\>[5]{}\Varid{load}\;\mathbin{:}\;\hidden{\sigma \;\tau }\;(\Varid{ctx}\;\mathbin{:}\;\Conid{EvalContext}\;\sigma \;\tau )\;(\Varid{c}\;\mathbin{:}\;\Conid{Closed}\;\sigma )\;\to \;\Conid{Decomposition}\;(\Varid{plug}\;\Varid{ctx}\;\Varid{c}){}\<[E]%
\\
\>[5]{}\Varid{load}\;\Varid{ctx}\;(\Conid{Closure}\;(\Conid{Lam}\;\Varid{body})\;\Varid{env})\;\mathrel{=}\;\Varid{unload}\;\Varid{ctx}\;\Varid{body}\;\Varid{env}{}\<[E]%
\\
\>[5]{}\Varid{load}\;\Varid{ctx}\;(\Conid{Closure}\;(\Conid{App}\;\Varid{f}\;\Varid{x})\;\Varid{env})\;\mathrel{=}\;\Conid{Decompose}\;(\Conid{Rapp}\;\Varid{f}\;\Varid{x}\;\Varid{env})\;\Varid{ctx}{}\<[E]%
\\
\>[5]{}\Varid{load}\;\Varid{ctx}\;(\Conid{Closure}\;(\Conid{Var}\;\Varid{i})\;\Varid{env})\;\mathrel{=}\;\Conid{Decompose}\;(\Conid{Lookup}\;\Varid{i}\;\Varid{env})\;\Varid{ctx}{}\<[E]%
\\
\>[5]{}\Varid{load}\;\Varid{ctx}\;(\Conid{Clapp}\;\Varid{f}\;\Varid{x})\;\mathrel{=}\;\Varid{load}\;(\Conid{ARG}\;\Varid{x}\;\Varid{ctx})\;\Varid{f}{}\<[E]%
\\[\blanklineskip]%
\>[5]{}\Varid{unload}\;\mathbin{:}\;\hidden{\sigma \;\tau \;\rho \;\Gamma }\;(\Varid{ctx}\;\mathbin{:}\;\Conid{EvalContext}\;(\sigma \;\Rightarrow\;\tau )\;\rho )\;(\Varid{body}\;\mathbin{:}\;\Conid{Term}\;(\Conid{Cons}\;\sigma \;\Gamma )\;\tau )\;(\Varid{env}\;\mathbin{:}\;\Conid{Env}\;\Gamma )\;{}\<[E]%
\\
\>[5]{}\hsindent{2}{}\<[7]%
\>[7]{}\to \;\Conid{Decomposition}\;(\Varid{plug}\;\Varid{ctx}\;(\Conid{Closure}\;(\Conid{Lam}\;\Varid{body})\;\Varid{env})){}\<[E]%
\\
\>[5]{}\Varid{unload}\;\Conid{MT}\;\Varid{body}\;\Varid{env}\;\mathrel{=}\;\Conid{Val}\;\Varid{body}\;\Varid{env}{}\<[E]%
\\
\>[5]{}\Varid{unload}\;(\Conid{ARG}\;\Varid{arg}\;\Varid{ctx})\;\Varid{body}\;\Varid{env}\;\mathrel{=}\;\Conid{Decompose}\;(\Conid{Beta}\;\Varid{body}\;\Varid{env}\;\Varid{arg})\;\Varid{ctx}{}\<[E]%
\ColumnHook
\end{hscode}\resethooks
The \ensuremath{\Varid{decompose}} function itself simply kicks off \ensuremath{\Varid{load}} with
an initially empty evaluation context.

\begin{hscode}\SaveRestoreHook
\column{B}{@{}>{\hspre}l<{\hspost}@{}}%
\column{3}{@{}>{\hspre}l<{\hspost}@{}}%
\column{E}{@{}>{\hspre}l<{\hspost}@{}}%
\>[3]{}\Varid{decompose}\;\mathbin{:}\;\hidden{\sigma }\;(\Varid{c}\;\mathbin{:}\;\Conid{Closed}\;\sigma )\;\to \;\Conid{Decomposition}\;\Varid{c}{}\<[E]%
\\
\>[3]{}\Varid{decompose}\;\Varid{c}\;\mathrel{=}\;\Varid{load}\;\Conid{MT}\;\Varid{c}{}\<[E]%
\ColumnHook
\end{hscode}\resethooks

To perform a single reduction step, we decompose a closed term. If
this yields a value, there is no further reduction to be done. If
decomposition yields a redex and evaluation context, we contract
the redex and plug the result back into the evaluation context:
\begin{hscode}\SaveRestoreHook
\column{B}{@{}>{\hspre}l<{\hspost}@{}}%
\column{3}{@{}>{\hspre}l<{\hspost}@{}}%
\column{E}{@{}>{\hspre}l<{\hspost}@{}}%
\>[3]{}\Varid{headReduce}\;\mathbin{:}\;\hidden{\sigma }\;\Conid{Closed}\;\sigma \;\to \;\Conid{Closed}\;\sigma {}\<[E]%
\\
\>[3]{}\Varid{headReduce}\;\Varid{c}\;\Keyword{with}\;\Varid{decompose}\;\Varid{c}{}\<[E]%
\\
\>[3]{}\Varid{headReduce}\;\lfloor \Conid{Closure}\;(\Conid{Lam}\;\Varid{body})\;\Varid{env}\rfloor\;\mid \;\Conid{Val}\;\Varid{body}\;\Varid{env}\;\mathrel{=}\;\Conid{Closure}\;(\Conid{Lam}\;\Varid{body})\;\Varid{env}{}\<[E]%
\\
\>[3]{}\Varid{headReduce}\;\lfloor \Varid{plug}\;\Varid{ctx}\;(\Varid{fromRedex}\;\Varid{redex})\rfloor\;\mid \;\Conid{Decompose}\;\Varid{redex}\;\Varid{ctx}\;\mathrel{=}\;\Varid{plug}\;\Varid{ctx}\;(\Varid{contract}\;\Varid{redex}){}\<[E]%
\ColumnHook
\end{hscode}\resethooks
Note that pattern matching on the \ensuremath{\Conid{Decomposition}} produces more
information about the term that has been decomposed. This is apparent
in the \emph{forced patterns}~\cite{norell}, \ensuremath{\lfloor \Conid{Closure}\;(\Conid{Lam}\;\Varid{body})\;\Varid{env}\rfloor} in the \ensuremath{\Conid{Val}} branch and \ensuremath{\lfloor \Varid{plug}\;\Varid{ctx}\;(\Varid{fromRedex}\;\Varid{redex})\rfloor} in
the \ensuremath{\Conid{Decompose}} branch, that appear on the left-hand side of the
function definition.

This completes our definition of a single head reduction step.

\section{Iterated head reduction}
\label{sec:iterated-head-reduction}
In the previous section we established how to perform a single
reduction step. Now it should be straightforward to define an
evaluation function by iteratively reducing by a single step until we
reach a value:

\begin{hscode}\SaveRestoreHook
\column{B}{@{}>{\hspre}l<{\hspost}@{}}%
\column{3}{@{}>{\hspre}l<{\hspost}@{}}%
\column{5}{@{}>{\hspre}l<{\hspost}@{}}%
\column{E}{@{}>{\hspre}l<{\hspost}@{}}%
\>[3]{}\Varid{evaluate}\;\mathbin{:}\;\hidden{\sigma }\;\Conid{Closed}\;\sigma \;\to \;\Conid{Value}\;\sigma {}\<[E]%
\\
\>[3]{}\Varid{evaluate}\;\Varid{c}\;\mathrel{=}\;\Varid{iterate}\;(\Varid{decompose}\;\Varid{c}){}\<[E]%
\\
\>[3]{}\hsindent{2}{}\<[5]%
\>[5]{}\Keyword{where}{}\<[E]%
\\
\>[3]{}\hsindent{2}{}\<[5]%
\>[5]{}\Varid{iterate}\;\mathbin{:}\;\hidden{\sigma }\;\Conid{Decomposition}\;\Varid{c}\;\to \;\Conid{Value}\;\sigma {}\<[E]%
\\
\>[3]{}\hsindent{2}{}\<[5]%
\>[5]{}\Varid{iterate}\;(\Conid{Val}\;\Varid{val}\;\Varid{p})\;\mathrel{=}\;\Conid{Val}\;\Varid{val}\;\Varid{p}{}\<[E]%
\\
\>[3]{}\hsindent{2}{}\<[5]%
\>[5]{}\Varid{iterate}\;(\Conid{Decompose}\;\Varid{r}\;\Varid{ctx})\;\mathrel{=}\;\Varid{iterate}\;(\Varid{decompose}\;(\Varid{plug}\;\Varid{ctx}\;(\Varid{contract}\;\Varid{r}))){}\<[E]%
\ColumnHook
\end{hscode}\resethooks

There is one problem with this definition: it is not structurally
recursive. It is rejected by Agda. Yet we know that the
simply typed lambda calculus is strongly normalizing---so iteratively
performing a single head reduction will always produce a value
eventually. How can we convince Agda of this fact?

The Bove-Capretta method is one technique to transform a definition
that is not structurally recursive into an equivalent definition that
is structurally recursive over a new
argument~\citep{bove-capretta}. Essentially, it does structural
recursion over the call graph of a function. In our case, we would
like to have an inhabitant of the following data type:

\begin{hscode}\SaveRestoreHook
\column{B}{@{}>{\hspre}l<{\hspost}@{}}%
\column{3}{@{}>{\hspre}l<{\hspost}@{}}%
\column{5}{@{}>{\hspre}l<{\hspost}@{}}%
\column{E}{@{}>{\hspre}l<{\hspost}@{}}%
\>[3]{}\Keyword{data}\;\Conid{Trace}\;\mathbin{:}\;\hidden{\sigma }\;\{\mskip1.5mu \Varid{c}\;\mathbin{:}\;\Conid{Closed}\;\sigma \mskip1.5mu\}\;\to \;\Conid{Decomposition}\;\Varid{c}\;\to \;\Conid{Set}\;\Keyword{where}{}\<[E]%
\\
\>[3]{}\hsindent{2}{}\<[5]%
\>[5]{}\Conid{Done}\;\mathbin{:}\;\hidden{\sigma \;\tau \;\Gamma }\;(\Varid{body}\;\mathbin{:}\;\Conid{Term}\;(\Conid{Cons}\;\sigma \;\Gamma )\;\tau )\;\to \;(\Varid{env}\;\mathbin{:}\;\Conid{Env}\;\Gamma )\;\to \;\Conid{Trace}\;(\Conid{Val}\;\Varid{body}\;\Varid{env}){}\<[E]%
\\
\>[3]{}\hsindent{2}{}\<[5]%
\>[5]{}\Conid{Step}\;\mathbin{:}\;\!\!\;\!\!\;\!\!\;\;\Conid{Trace}\;(\Varid{decompose}\;(\Varid{plug}\;\Varid{ctx}\;(\Varid{contract}\;\Varid{r})))\;\to \;\Conid{Trace}\;(\Conid{Decompose}\;\Varid{r}\;\Varid{ctx}){}\<[E]%
\ColumnHook
\end{hscode}\resethooks
We could then define the \ensuremath{\Varid{iterate}} function by structural induction
over the trace:
\begin{hscode}\SaveRestoreHook
\column{B}{@{}>{\hspre}l<{\hspost}@{}}%
\column{3}{@{}>{\hspre}l<{\hspost}@{}}%
\column{E}{@{}>{\hspre}l<{\hspost}@{}}%
\>[3]{}\Varid{iterate}\;\mathbin{:}\;\hidden{\sigma \;\mathbin{:}\;\Conid{Ty}}\;\{\mskip1.5mu \Varid{c}\;\mathbin{:}\;\Conid{Closed}\;\sigma \mskip1.5mu\}\;\to \;(\Varid{d}\;\mathbin{:}\;\Conid{Decomposition}\;\Varid{c})\;\to \;\Conid{Trace}\;\Varid{d}\;\to \;\Conid{Value}\;\sigma {}\<[E]%
\\
\>[3]{}\Varid{iterate}\;(\Conid{Val}\;\Varid{body}\;\Varid{env})\;(\Conid{Done}\;\lfloor \Varid{body}\rfloor\;\lfloor \Varid{env}\rfloor)\;\mathrel{=}\;\Conid{Val}\;(\Conid{Closure}\;(\Conid{Lam}\;\Varid{body})\;\Varid{env})\;\Varid{unit}{}\<[E]%
\\
\>[3]{}\Varid{iterate}\;(\Conid{Decompose}\;\Varid{r}\;\Varid{ctx})\;(\Conid{Step}\;\Varid{step})\;\mathrel{=}\;\Varid{iterate}\;(\Varid{decompose}\;(\Varid{plug}\;\Varid{ctx}\;(\Varid{contract}\;\Varid{r})))\;\Varid{step}{}\<[E]%
\ColumnHook
\end{hscode}\resethooks
Although this definition does pass Agda's termination checker, the
question remains how to provide the required \ensuremath{\Conid{Trace}} argument to
our \ensuremath{\Varid{iterate}} function. That is we would like to define a function of
type:
\begin{hscode}\SaveRestoreHook
\column{B}{@{}>{\hspre}l<{\hspost}@{}}%
\column{3}{@{}>{\hspre}l<{\hspost}@{}}%
\column{E}{@{}>{\hspre}l<{\hspost}@{}}%
\>[3]{}\hidden{\sigma }\;(\Varid{t}\;\mathbin{:}\;\Conid{Closed}\;\sigma )\;\to \;\Conid{Trace}\;\Varid{t}{}\<[E]%
\ColumnHook
\end{hscode}\resethooks
A straightforward attempt to define such a function fails
immediately. Instead, we need to define the following \emph{logical
  relation} that strengthens our induction hypothesis:

\begin{hscode}\SaveRestoreHook
\column{B}{@{}>{\hspre}l<{\hspost}@{}}%
\column{3}{@{}>{\hspre}l<{\hspost}@{}}%
\column{27}{@{}>{\hspre}l<{\hspost}@{}}%
\column{33}{@{}>{\hspre}l<{\hspost}@{}}%
\column{E}{@{}>{\hspre}l<{\hspost}@{}}%
\>[3]{}\Conid{Reducible}\;\mathbin{:}\;\{\mskip1.5mu \sigma \;\mathbin{:}\;\Conid{Ty}\mskip1.5mu\}\;\to \;(\Varid{t}\;\mathbin{:}\;\Conid{Closed}\;\sigma )\;\to \;\Conid{Set}{}\<[E]%
\\
\>[3]{}\Conid{Reducible}\;\{\mskip1.5mu \Conid{O}\mskip1.5mu\}\;\Varid{t}\;\mathrel{=}\;\Conid{Trace}\;(\Varid{decompose}\;\Varid{t}){}\<[E]%
\\
\>[3]{}\Conid{Reducible}\;\{\mskip1.5mu \sigma \;\Rightarrow\;\tau \mskip1.5mu\}\;\Varid{t}\;\mathrel{=}\;{}\<[27]%
\>[27]{}\Conid{Pair}\;{}\<[33]%
\>[33]{}(\Conid{Trace}\;(\Varid{decompose}\;\Varid{t}))\;{}\<[E]%
\\
\>[33]{}((\Varid{x}\;\mathbin{:}\;\Conid{Closed}\;\sigma )\;\to \;\Conid{Reducible}\;\Varid{x}\;\to \;\Conid{Reducible}\;(\Conid{Clapp}\;\Varid{t}\;\Varid{x})){}\<[E]%
\\[\blanklineskip]%
\>[3]{}\Conid{ReducibleEnv}\;\mathbin{:}\;\hidden{\Gamma }\;\Conid{Env}\;\Gamma \;\to \;\Conid{Set}{}\<[E]%
\\
\>[3]{}\Conid{ReducibleEnv}\;\Conid{Nil}\;\mathrel{=}\;\Conid{Unit}{}\<[E]%
\\
\>[3]{}\Conid{ReducibleEnv}\;(\Varid{x}\;\Varid{·}\;\Varid{env})\;\mathrel{=}\;\Conid{Pair}\;(\Conid{Reducible}\;\Varid{x})\;(\Conid{ReducibleEnv}\;\Varid{env}){}\<[E]%
\ColumnHook
\end{hscode}\resethooks

To prove that all closed terms are reducible, we follow the proof
sketched by Coquand~\cite{coquand} and prove the following two lemmas.

\begin{hscode}\SaveRestoreHook
\column{B}{@{}>{\hspre}l<{\hspost}@{}}%
\column{3}{@{}>{\hspre}l<{\hspost}@{}}%
\column{E}{@{}>{\hspre}l<{\hspost}@{}}%
\>[3]{}\Varid{lemma1}\;\mathbin{:}\;\hidden{\sigma }\;(\Varid{c}\;\mathbin{:}\;\Conid{Closed}\;\sigma )\;\to \;\Conid{Reducible}\;(\Varid{headReduce}\;\Varid{c})\;\to \;\Conid{Reducible}\;\Varid{c}{}\<[E]%
\\[\blanklineskip]%
\>[3]{}\Varid{lemma2}\;\mathbin{:}\;\hidden{\Gamma \;\sigma }\;(\Varid{t}\;\mathbin{:}\;\Conid{Term}\;\Gamma \;\sigma )\;(\Varid{env}\;\mathbin{:}\;\Conid{Env}\;\Gamma )\;\to \;\Conid{ReducibleEnv}\;\Varid{env}\;\to \;\Conid{Reducible}\;(\Conid{Closure}\;\Varid{t}\;\Varid{env}){}\<[E]%
\ColumnHook
\end{hscode}\resethooks
The proof of \ensuremath{\Varid{lemma2}} performs induction on the term \ensuremath{\Varid{t}}. In each of
the branches, we appeal to \ensuremath{\Varid{lemma1}} in order to prove that \ensuremath{\Conid{Closure}\;\Varid{t}\;\Varid{env}} is also reducible. The proof of \ensuremath{\Varid{lemma1}} is done by induction on
\ensuremath{\sigma } and \ensuremath{\Varid{c}}. The only difficult case is that for closed applications,
\ensuremath{\Conid{Clapp}\;\Varid{f}\;\Varid{x}}. In that branch, we need to show that \ensuremath{\Conid{Clapp}\;(\Varid{headReduce}\;(\Conid{Clapp}\;\Varid{f}\;\Varid{x}))\;\Varid{y}} is equal to \ensuremath{\Varid{headReduce}\;(\Conid{Clapp}\;(\Conid{Clapp}\;\Varid{f}\;\Varid{x})\;\Varid{y})}.

To prove the desired equality we observe that if decomposing \ensuremath{\Conid{Clapp}\;\Varid{f}\;\Varid{x}} yields a redex \ensuremath{\Varid{r}} and evaluation context \ensuremath{\Varid{ctx}}, then the
decomposition of \ensuremath{\Conid{Clapp}\;(\Conid{Clapp}\;\Varid{f}\;\Varid{x})\;\Varid{y}} must yield the same redex with
the evaluation context obtained by adding \ensuremath{\Varid{y}} to the end of \ensuremath{\Varid{ctx}}. To
complete the proof we define an auxiliary `backwards view' on
evaluation contexts that states that every evaluation context is
either empty or arises by adding a closed term to the end of an
evaluation context. Using this view, the required equality is
easy to prove.

Using \ensuremath{\Varid{lemma1}} and \ensuremath{\Varid{lemma2}}, we can prove our main theorem: every
closed term is reducible. To do so, we define the following two
mutually recursive theorems:
\begin{hscode}\SaveRestoreHook
\column{B}{@{}>{\hspre}l<{\hspost}@{}}%
\column{3}{@{}>{\hspre}l<{\hspost}@{}}%
\column{5}{@{}>{\hspre}l<{\hspost}@{}}%
\column{E}{@{}>{\hspre}l<{\hspost}@{}}%
\>[3]{}\Keyword{mutual}{}\<[E]%
\\
\>[3]{}\hsindent{2}{}\<[5]%
\>[5]{}\Varid{theorem}\;\mathbin{:}\;\hidden{\sigma }\;(\Varid{c}\;\mathbin{:}\;\Conid{Closed}\;\sigma )\;\to \;\Conid{Reducible}\;\Varid{c}{}\<[E]%
\\
\>[3]{}\hsindent{2}{}\<[5]%
\>[5]{}\Varid{theorem}\;(\Conid{Closure}\;\Varid{t}\;\Varid{env})\;\mathrel{=}\;\Varid{lemma2}\;\Varid{t}\;\Varid{env}\;(\Varid{envTheorem}\;\Varid{env}){}\<[E]%
\\
\>[3]{}\hsindent{2}{}\<[5]%
\>[5]{}\Varid{theorem}\;(\Conid{Clapp}\;\Varid{f}\;\Varid{x})\;\mathrel{=}\;\Varid{snd}\;(\Varid{theorem}\;\Varid{f})\;\Varid{x}\;(\Varid{theorem}\;\Varid{x}){}\<[E]%
\\[\blanklineskip]%
\>[3]{}\hsindent{2}{}\<[5]%
\>[5]{}\Varid{envTheorem}\;\mathbin{:}\;\hidden{\Gamma }\;(\Varid{env}\;\mathbin{:}\;\Conid{Env}\;\Gamma )\;\to \;\Conid{ReducibleEnv}\;\Varid{env}{}\<[E]%
\\
\>[3]{}\hsindent{2}{}\<[5]%
\>[5]{}\Varid{envTheorem}\;\Conid{Nil}\;\mathrel{=}\;\Varid{unit}{}\<[E]%
\\
\>[3]{}\hsindent{2}{}\<[5]%
\>[5]{}\Varid{envTheorem}\;(\Varid{t}\;\Varid{·}\;\Varid{ts})\;\mathrel{=}\;(\Varid{theorem}\;\Varid{t},\Varid{envTheorem}\;\Varid{ts}){}\<[E]%
\ColumnHook
\end{hscode}\resethooks
To prove that every closure is reducible, we appeal to \ensuremath{\Varid{lemma2}} and
prove that every closed term in the environment is also reducible. The
proof that every closed application is reducible recurses over both
arguments \ensuremath{\Varid{f}} and \ensuremath{\Varid{x}}. The recursive call to \ensuremath{\Varid{f}} yields a pair of a
trace and a function of type:
\begin{hscode}\SaveRestoreHook
\column{B}{@{}>{\hspre}l<{\hspost}@{}}%
\column{3}{@{}>{\hspre}l<{\hspost}@{}}%
\column{E}{@{}>{\hspre}l<{\hspost}@{}}%
\>[3]{}((\Varid{x}\;\mathbin{:}\;\Conid{Closed}\;\sigma )\;\to \;\Conid{Reducible}\;\Varid{x}\;\to \;\Conid{Reducible}\;(\Conid{Clapp}\;\Varid{f}\;\Varid{x})){}\<[E]%
\ColumnHook
\end{hscode}\resethooks
Applying this function to \ensuremath{\Varid{x}} and \ensuremath{\Varid{theorem}\;\Varid{x}}, yields the desired
proof.

One important corollary of our theorem is that for every closed term
\ensuremath{\Varid{c}}, we can compute an evaluation trace of \ensuremath{\Varid{c}}:
\begin{hscode}\SaveRestoreHook
\column{B}{@{}>{\hspre}l<{\hspost}@{}}%
\column{3}{@{}>{\hspre}l<{\hspost}@{}}%
\column{E}{@{}>{\hspre}l<{\hspost}@{}}%
\>[3]{}\Varid{termination}\;\mathbin{:}\;\{\mskip1.5mu \sigma \;\mathbin{:}\;\Conid{Ty}\mskip1.5mu\}\;\to \;(\Varid{c}\;\mathbin{:}\;\Conid{Closed}\;\sigma )\;\to \;\Conid{Trace}\;(\Varid{decompose}\;\Varid{c}){}\<[E]%
\\
\>[3]{}\Varid{termination}\;\{\mskip1.5mu \Conid{O}\mskip1.5mu\}\;\Varid{c}\;\mathrel{=}\;\Varid{theorem}\;\Varid{c}{}\<[E]%
\\
\>[3]{}\Varid{termination}\;\{\mskip1.5mu \sigma \;\Rightarrow\;\tau \mskip1.5mu\}\;\Varid{c}\;\mathrel{=}\;\Varid{fst}\;(\Varid{theorem}\;\Varid{c}){}\<[E]%
\ColumnHook
\end{hscode}\resethooks
Now we can finally complete the definition of our small step
evaluation function:
\begin{hscode}\SaveRestoreHook
\column{B}{@{}>{\hspre}l<{\hspost}@{}}%
\column{3}{@{}>{\hspre}l<{\hspost}@{}}%
\column{E}{@{}>{\hspre}l<{\hspost}@{}}%
\>[3]{}\Varid{evaluate}\;\mathbin{:}\;\hidden{\sigma }\;\Conid{Closed}\;\sigma \;\to \;\Conid{Value}\;\sigma {}\<[E]%
\\
\>[3]{}\Varid{evaluate}\;\Varid{t}\;\mathrel{=}\;\Varid{iterate}\;(\Varid{decompose}\;\Varid{t})\;(\Varid{termination}\;\Varid{t}){}\<[E]%
\ColumnHook
\end{hscode}\resethooks
The \ensuremath{\Varid{evaluate}} function iteratively performs a single step of head
reduction, performing structural induction over the trace that we
compute using the reducibility proof sketched above.

\section{Refocusing}
\label{sec:refocusing}

The small step evaluator presented in the previous section repeatedly
decomposes a closed term into an evaluation context and a redex,
contracts the redex, and plugs the contractum back into the evaluation
context. Before transforming this evaluator into the Krivine machine,
we will show how to apply the refocusing transformation to produce a
\emph{small-step abstract machine}~\cite{danvy:small-step}. This
small-step abstract machine forms a convenient halfway point between
the small step evaluator and the Krivine machine.

The key idea of refocusing is to compose the plugging and
decomposition steps into a single \ensuremath{\Varid{refocus}} operation. Instead of
repeatedly plugging and decomposing, the \ensuremath{\Varid{refocus}} function navigates
directly to the next redex, if it exists:
\begin{hscode}\SaveRestoreHook
\column{B}{@{}>{\hspre}l<{\hspost}@{}}%
\column{3}{@{}>{\hspre}l<{\hspost}@{}}%
\column{E}{@{}>{\hspre}l<{\hspost}@{}}%
\>[3]{}\Varid{refocus}\;\mathbin{:}\;\hidden{\sigma \;\tau }\;(\Varid{ctx}\;\mathbin{:}\;\Conid{EvalContext}\;\sigma \;\tau )\;(\Varid{c}\;\mathbin{:}\;\Conid{Closed}\;\sigma )\;\to \;\Conid{Decomposition}\;(\Varid{plug}\;\Varid{ctx}\;\Varid{c}){}\<[E]%
\\
\>[3]{}\Varid{refocus}\;\Conid{MT}\;(\Conid{Closure}\;(\Conid{Lam}\;\Varid{body})\;\Varid{env})\;\mathrel{=}\;\Conid{Val}\;\Varid{body}\;\Varid{env}{}\<[E]%
\\
\>[3]{}\Varid{refocus}\;(\Conid{ARG}\;\Varid{x}\;\Varid{ctx})\;(\Conid{Closure}\;(\Conid{Lam}\;\Varid{body})\;\Varid{env})\;\mathrel{=}\;\Conid{Decompose}\;(\Conid{Beta}\;\Varid{body}\;\Varid{env}\;\Varid{x})\;\Varid{ctx}{}\<[E]%
\\
\>[3]{}\Varid{refocus}\;\Varid{ctx}\;(\Conid{Closure}\;(\Conid{Var}\;\Varid{i})\;\Varid{env})\;\mathrel{=}\;\Conid{Decompose}\;(\Conid{Lookup}\;\Varid{i}\;\Varid{env})\;\Varid{ctx}{}\<[E]%
\\
\>[3]{}\Varid{refocus}\;\Varid{ctx}\;(\Conid{Closure}\;(\Conid{App}\;\Varid{f}\;\Varid{x})\;\Varid{env})\;\mathrel{=}\;\Conid{Decompose}\;(\Conid{Rapp}\;\Varid{f}\;\Varid{x}\;\Varid{env})\;\Varid{ctx}{}\<[E]%
\\
\>[3]{}\Varid{refocus}\;\Varid{ctx}\;(\Conid{Clapp}\;\Varid{f}\;\Varid{x})\;\mathrel{=}\;\Varid{refocus}\;(\Conid{ARG}\;\Varid{x}\;\Varid{ctx})\;\Varid{f}{}\<[E]%
\ColumnHook
\end{hscode}\resethooks
We can formalize this intuition about the behaviour of refocusing by
proving the following lemma:
\begin{hscode}\SaveRestoreHook
\column{B}{@{}>{\hspre}l<{\hspost}@{}}%
\column{3}{@{}>{\hspre}l<{\hspost}@{}}%
\column{6}{@{}>{\hspre}l<{\hspost}@{}}%
\column{E}{@{}>{\hspre}l<{\hspost}@{}}%
\>[3]{}\Varid{refocusCorrect}\;\mathbin{:}\;\hidden{\tau \;\sigma }\;(\Varid{ctx}\;\mathbin{:}\;\Conid{EvalContext}\;\sigma \;\tau )\;(\Varid{c}\;\mathbin{:}\;\Conid{Closed}\;\sigma )\;\to \;{}\<[E]%
\\
\>[3]{}\hsindent{3}{}\<[6]%
\>[6]{}\Varid{refocus}\;\Varid{ctx}\;\Varid{c}\;\equiv \;\Varid{decompose}\;(\Varid{plug}\;\Varid{ctx}\;\Varid{c}){}\<[E]%
\ColumnHook
\end{hscode}\resethooks
The proof by induction on \ensuremath{\Varid{ctx}} and \ensuremath{\Varid{c}} relies on an easy lemma:
\begin{hscode}\SaveRestoreHook
\column{B}{@{}>{\hspre}l<{\hspost}@{}}%
\column{3}{@{}>{\hspre}l<{\hspost}@{}}%
\column{5}{@{}>{\hspre}l<{\hspost}@{}}%
\column{E}{@{}>{\hspre}l<{\hspost}@{}}%
\>[3]{}\Varid{decomposePlug}\;\mathbin{:}\;\hidden{\sigma \;\tau }\;(\Varid{ctx}\;\mathbin{:}\;\Conid{EvalContext}\;\sigma \;\tau )\;(\Varid{c}\;\mathbin{:}\;\Conid{Closed}\;\sigma )\;\to \;{}\<[E]%
\\
\>[3]{}\hsindent{2}{}\<[5]%
\>[5]{}\Varid{decompose}\;(\Varid{plug}\;\Varid{ctx}\;\Varid{c})\;\equiv \;\Varid{load}\;\Varid{ctx}\;\Varid{c}{}\<[E]%
\ColumnHook
\end{hscode}\resethooks
The proof of the \ensuremath{\Varid{decomposePlug}} lemma proceeds by simple induction on the
evaluation context.

To rewrite our evaluator to use the \ensuremath{\Varid{refocus}} operation, we will need to
adapt the \ensuremath{\Conid{Trace}} data type from the previous section. Iterated
recursive calls will no longer call \ensuremath{\Varid{decompose}} and \ensuremath{\Varid{plug}}, but
instead navigate to the next redex using the \ensuremath{\Varid{refocus}} function. The
new \ensuremath{\Conid{Trace}} data type reflects just that:

\begin{hscode}\SaveRestoreHook
\column{B}{@{}>{\hspre}l<{\hspost}@{}}%
\column{3}{@{}>{\hspre}l<{\hspost}@{}}%
\column{6}{@{}>{\hspre}l<{\hspost}@{}}%
\column{130}{@{}>{\hspre}l<{\hspost}@{}}%
\column{E}{@{}>{\hspre}l<{\hspost}@{}}%
\>[3]{}\Keyword{data}\;\Conid{Trace}\;\mathbin{:}\;\!\!\;\!\!\;\;\Conid{Decomposition}\;\Varid{c}\;\to \;\Conid{Set}\;\Keyword{where}{}\<[E]%
\\
\>[3]{}\hsindent{3}{}\<[6]%
\>[6]{}\Conid{Done}\;\mathbin{:}\;\hidden{\Gamma \;\sigma \;\tau }\;(\Varid{body}\;\mathbin{:}\;\Conid{Term}\;(\Conid{Cons}\;\sigma \;\Gamma )\;\tau )\;\to \;(\Varid{env}\;\mathbin{:}\;\Conid{Env}\;\Gamma )\;\to \;\Conid{Trace}\;(\Conid{Val}\;\Varid{body}\;\Varid{env}){}\<[E]%
\\
\>[3]{}\hsindent{3}{}\<[6]%
\>[6]{}\Conid{Step}\;\mathbin{:}\;\hidden{\sigma \;\tau }\;\!\!\;\!\!\;\;\Conid{Trace}\;(\Varid{refocus}\;\Varid{ctx}\;(\Varid{contract}\;\Varid{r}))\;\to \;\Conid{Trace}\;{}\<[130]%
\>[130]{}(\Conid{Decompose}\;\Varid{r}\;\Varid{ctx}){}\<[E]%
\ColumnHook
\end{hscode}\resethooks

To prove that this new \ensuremath{\Conid{Trace}} data type is inhabited, we call the
\ensuremath{\Varid{termination}} lemma from the previous section. Using the
\ensuremath{\Varid{refocusCorrect}} lemma, we perform induction on the \ensuremath{\Conid{Trace}} data type
from the previous section to construct a witness of termination. All
this is done by the following \ensuremath{\Varid{termination}} function:

\begin{hscode}\SaveRestoreHook
\column{B}{@{}>{\hspre}l<{\hspost}@{}}%
\column{3}{@{}>{\hspre}l<{\hspost}@{}}%
\column{E}{@{}>{\hspre}l<{\hspost}@{}}%
\>[3]{}\Varid{termination}\;\mathbin{:}\;\hidden{\sigma }\;(\Varid{c}\;\mathbin{:}\;\Conid{Closed}\;\sigma )\;\to \;\Conid{Trace}\;(\Varid{refocus}\;\Conid{MT}\;\Varid{c}){}\<[E]%
\ColumnHook
\end{hscode}\resethooks

The definition of our evaluator is now straightforward. The \ensuremath{\Varid{iterate}}
function repeatedly refocuses and contracts until a value has been
reached:
\begin{hscode}\SaveRestoreHook
\column{B}{@{}>{\hspre}l<{\hspost}@{}}%
\column{3}{@{}>{\hspre}l<{\hspost}@{}}%
\column{E}{@{}>{\hspre}l<{\hspost}@{}}%
\>[3]{}\Varid{iterate}\;\mathbin{:}\;\hidden{\sigma }\;\!\!\;\;(\Varid{d}\;\mathbin{:}\;\Conid{Decomposition}\;\Varid{c})\;\to \;\Conid{Trace}\;\Varid{d}\;\to \;\Conid{Value}\;\sigma {}\<[E]%
\\
\>[3]{}\Varid{iterate}\;(\Conid{Val}\;\Varid{body}\;\Varid{env})\;(\Conid{Done}\;\lfloor \Varid{body}\rfloor\;\lfloor \Varid{env}\rfloor)\;\mathrel{=}\;\Conid{Val}\;(\Conid{Closure}\;(\Conid{Lam}\;\Varid{body})\;\Varid{env})\;\Varid{unit}{}\<[E]%
\\
\>[3]{}\Varid{iterate}\;(\Conid{Decompose}\;\Varid{r}\;\Varid{ctx})\;(\Conid{Step}\;\Varid{step})\;\mathrel{=}\;\Varid{iterate}\;(\Varid{refocus}\;\Varid{ctx}\;(\Varid{contract}\;\Varid{r}))\;\Varid{step}{}\<[E]%
\\[\blanklineskip]%
\>[3]{}\Varid{evaluate}\;\mathbin{:}\;\hidden{\sigma }\;\Conid{Closed}\;\sigma \;\to \;\Conid{Value}\;\sigma {}\<[E]%
\\
\>[3]{}\Varid{evaluate}\;\Varid{c}\;\mathrel{=}\;\Varid{iterate}\;(\Varid{refocus}\;\Conid{MT}\;\Varid{c})\;(\Varid{termination}\;\Varid{c}){}\<[E]%
\ColumnHook
\end{hscode}\resethooks
The \ensuremath{\Varid{evaluate}} function kicks off the \ensuremath{\Varid{iterate}} function with
an empty evaluation context and a proof of termination.

Finally, we can also show that our new evaluator behaves the same as
the evaluation function presented in the previous section. To do so,
we prove the following lemma by induction on the decomposition of \ensuremath{\Varid{t}}:

\begin{hscode}\SaveRestoreHook
\column{B}{@{}>{\hspre}l<{\hspost}@{}}%
\column{3}{@{}>{\hspre}l<{\hspost}@{}}%
\column{5}{@{}>{\hspre}l<{\hspost}@{}}%
\column{E}{@{}>{\hspre}l<{\hspost}@{}}%
\>[3]{}\Varid{correctness}\;\mathbin{:}\;\hidden{\sigma }\;\{\mskip1.5mu \Varid{t}\;\mathbin{:}\;\Conid{Closed}\;\sigma \mskip1.5mu\}\;\to \;{}\<[E]%
\\
\>[3]{}\hsindent{2}{}\<[5]%
\>[5]{}(\Varid{trace}\;\mathbin{:}\;\Conid{Trace}\;(\Varid{refocus}\;\Conid{MT}\;\Varid{t}))\;\to \;(\Varid{trace'}\;\mathbin{:}\;\Conid{Section4.Trace}\;(\Varid{decompose}\;\Varid{t}))\;\to \;{}\<[E]%
\\
\>[3]{}\hsindent{2}{}\<[5]%
\>[5]{}\Varid{iterate}\;(\Varid{refocus}\;\Conid{MT}\;\Varid{t})\;\Varid{trace}\;\equiv \;\Conid{Section4.iterate}\;(\Varid{decompose}\;\Varid{t})\;\Varid{trace'}{}\<[E]%
\ColumnHook
\end{hscode}\resethooks
An important corollary of this \ensuremath{\Varid{correctness}} property is that our new
evaluation function behaves identically to the \ensuremath{\Varid{evaluate}} function
from the previous section:

\begin{hscode}\SaveRestoreHook
\column{B}{@{}>{\hspre}l<{\hspost}@{}}%
\column{3}{@{}>{\hspre}l<{\hspost}@{}}%
\column{E}{@{}>{\hspre}l<{\hspost}@{}}%
\>[3]{}\Varid{corollary}\;\mathbin{:}\;\hidden{\sigma }\;(\Varid{t}\;\mathbin{:}\;\Conid{Closed}\;\sigma )\;\to \;\Varid{evaluate}\;\Varid{t}\;\equiv \;\Conid{Section4.evaluate}\;\Varid{t}{}\<[E]%
\\
\>[3]{}\Varid{corollary}\;\Varid{t}\;\mathrel{=}\;\Varid{correctness}\;(\Varid{termination}\;\Varid{t})\;(\Conid{Section4.termination}\;\Varid{t}){}\<[E]%
\ColumnHook
\end{hscode}\resethooks

This completes the definition and verification of the evaluator that
arises by applying the refocusing transformation on the small step
evaluator from Section~\ref{sec:iterated-head-reduction}.

\section{The Krivine machine}
\label{sec:krivine}

In this section we will derive the Krivine machine from the evaluation
function we saw previously. To complete our derivation, we perform a
few further program transformations on the previous evaluation
function. 

We start by inlining the \ensuremath{\Varid{iterate}} function, making our \ensuremath{\Varid{refocus}}
function recursive. Furthermore, the \ensuremath{\Varid{evaluate}} function in the
previous section mapped \ensuremath{\Conid{App}} terms into closed \ensuremath{\Conid{Clapp}} terms, and
subsequently evaluated the first argument of the resulting \ensuremath{\Conid{Clapp}}
constructor, adding the second argument to the evaluation context. In
this section, we will combine these two steps into a single
transition---a transformation sometimes referred to as
\emph{compressing corridor transitions}~\cite{danvy-afp}. As a result,
we will no longer add closed applications to the environment or
evaluation context. We introduce the following predicates enforcing
the absence of \ensuremath{\Conid{Clapp}} constructors on closed terms, environments, and
evaluation contexts respectively:

\begin{hscode}\SaveRestoreHook
\column{B}{@{}>{\hspre}l<{\hspost}@{}}%
\column{3}{@{}>{\hspre}l<{\hspost}@{}}%
\column{5}{@{}>{\hspre}l<{\hspost}@{}}%
\column{E}{@{}>{\hspre}l<{\hspost}@{}}%
\>[3]{}\Keyword{mutual}{}\<[E]%
\\
\>[3]{}\hsindent{2}{}\<[5]%
\>[5]{}\Varid{isValidClosure}\;\mathbin{:}\;\hidden{\sigma }\;\Conid{Closed}\;\sigma \;\to \;\Conid{Set}{}\<[E]%
\\
\>[3]{}\hsindent{2}{}\<[5]%
\>[5]{}\Varid{isValidClosure}\;(\Conid{Closure}\;\Varid{t}\;\Varid{env})\;\mathrel{=}\;\Varid{isValidEnv}\;\Varid{env}{}\<[E]%
\\
\>[3]{}\hsindent{2}{}\<[5]%
\>[5]{}\Varid{isValidClosure}\;(\Conid{Clapp}\;\Varid{f}\;\Varid{x})\;\mathrel{=}\;\Conid{Empty}{}\<[E]%
\\[\blanklineskip]%
\>[3]{}\hsindent{2}{}\<[5]%
\>[5]{}\Varid{isValidEnv}\;\mathbin{:}\;\hidden{\Delta }\;\Conid{Env}\;\Delta \;\to \;\Conid{Set}{}\<[E]%
\\
\>[3]{}\hsindent{2}{}\<[5]%
\>[5]{}\Varid{isValidEnv}\;\Conid{Nil}\;\mathrel{=}\;\Conid{Unit}{}\<[E]%
\\
\>[3]{}\hsindent{2}{}\<[5]%
\>[5]{}\Varid{isValidEnv}\;(\Varid{c}\;\Varid{·}\;\Varid{env})\;\mathrel{=}\;\Conid{Pair}\;(\Varid{isValidClosure}\;\Varid{c})\;(\Varid{isValidEnv}\;\Varid{env}){}\<[E]%
\\[\blanklineskip]%
\>[3]{}\Varid{isValidContext}\;\mathbin{:}\;\hidden{\sigma \;\tau }\;\Conid{EvalContext}\;\sigma \;\tau \;\to \;\Conid{Set}{}\<[E]%
\\
\>[3]{}\Varid{isValidContext}\;\Conid{MT}\;\mathrel{=}\;\Conid{Unit}{}\<[E]%
\\
\>[3]{}\Varid{isValidContext}\;(\Conid{ARG}\;(\Conid{Closure}\;\Varid{t}\;\Varid{env})\;\Varid{ctx})\;\mathrel{=}\;\Conid{Pair}\;(\Varid{isValidEnv}\;\Varid{env})\;(\Varid{isValidContext}\;\Varid{ctx}){}\<[E]%
\\
\>[3]{}\Varid{isValidContext}\;(\Conid{ARG}\;(\Conid{Clapp}\;\Varid{f}\;\Varid{x})\;\Varid{env})\;\mathrel{=}\;\Conid{Empty}{}\<[E]%
\ColumnHook
\end{hscode}\resethooks

\pagebreak

Given that the only valid closed terms are closures, we can define
functions that project the underlying environment and term from any
valid closed term:

\begin{hscode}\SaveRestoreHook
\column{B}{@{}>{\hspre}l<{\hspost}@{}}%
\column{3}{@{}>{\hspre}l<{\hspost}@{}}%
\column{E}{@{}>{\hspre}l<{\hspost}@{}}%
\>[3]{}\Varid{getContext}\;\mathbin{:}\;\hidden{\sigma }\;\Conid{Exists}\;(\Conid{Closed}\;\sigma )\;\Varid{isValidClosure}\;\to \;\Conid{Context}{}\<[E]%
\\
\>[3]{}\Varid{getContext}\;(\Conid{Witness}\;(\Conid{Closure}\;\{\mskip1.5mu \Gamma \mskip1.5mu\}\;\Varid{t}\;\Varid{env})\;\anonymous )\;\mathrel{=}\;\Gamma {}\<[E]%
\\
\>[3]{}\Varid{getContext}\;(\Conid{Witness}\;(\Conid{Clapp}\;\Varid{f}\;\Varid{x})\;()){}\<[E]%
\\[\blanklineskip]%
\>[3]{}\Varid{getEnv}\;\mathbin{:}\;\hidden{\sigma }\;(\Varid{c}\;\mathbin{:}\;\Conid{Exists}\;(\Conid{Closed}\;\sigma )\;\Varid{isValidClosure})\;\to \;\Conid{Env}\;(\Varid{getContext}\;\Varid{c}){}\<[E]%
\\
\>[3]{}\Varid{getEnv}\;(\Conid{Witness}\;(\Conid{Closure}\;\Varid{t}\;\Varid{env})\;\Varid{p})\;\mathrel{=}\;\Varid{env}{}\<[E]%
\\
\>[3]{}\Varid{getEnv}\;(\Conid{Witness}\;(\Conid{Clapp}\;\Varid{f}\;\Varid{x})\;()){}\<[E]%
\\[\blanklineskip]%
\>[3]{}\Varid{getTerm}\;\mathbin{:}\;\hidden{\sigma }\;(\Varid{c}\;\mathbin{:}\;\Conid{Exists}\;(\Conid{Closed}\;\sigma )\;\Varid{isValidClosure})\;\to \;\Conid{Term}\;(\Varid{getContext}\;\Varid{c})\;\sigma {}\<[E]%
\\
\>[3]{}\Varid{getTerm}\;(\Conid{Witness}\;(\Conid{Closure}\;\Varid{t}\;\Varid{env})\;\Varid{p})\;\mathrel{=}\;\Varid{t}{}\<[E]%
\\
\>[3]{}\Varid{getTerm}\;(\Conid{Witness}\;(\Conid{Clapp}\;\Varid{f}\;\Varid{x})\;()){}\<[E]%
\ColumnHook
\end{hscode}\resethooks

Finally, we can define a new \ensuremath{\Varid{lookup}} operation that guarantees that
looking up a variable in a valid environment will always return a
closure:
\begin{hscode}\SaveRestoreHook
\column{B}{@{}>{\hspre}l<{\hspost}@{}}%
\column{3}{@{}>{\hspre}l<{\hspost}@{}}%
\column{5}{@{}>{\hspre}l<{\hspost}@{}}%
\column{E}{@{}>{\hspre}l<{\hspost}@{}}%
\>[3]{}\Varid{lookup}\;\mathbin{:}\;\hidden{\sigma \;\Gamma }\;\Conid{Ref}\;\Gamma \;\sigma \;\to \;(\Varid{env}\;\mathbin{:}\;\Conid{Env}\;\Gamma )\;\to \;\Varid{isValidEnv}\;\Varid{env}\;\to \;{}\<[E]%
\\
\>[3]{}\hsindent{2}{}\<[5]%
\>[5]{}\Conid{Exists}\;(\Conid{Closed}\;\sigma )\;\Varid{isValidClosure}{}\<[E]%
\\
\>[3]{}\Varid{lookup}\;\Conid{Top}\;(\Conid{Closure}\;\Varid{t}\;\Varid{env}\;\Varid{·}\;\anonymous )\;(\Varid{p1},\Varid{p2})\;\mathrel{=}\;\Conid{Witness}\;(\Conid{Closure}\;\Varid{t}\;\Varid{env})\;\Varid{p1}{}\<[E]%
\\
\>[3]{}\Varid{lookup}\;\Conid{Top}\;(\Conid{Clapp}\;\anonymous \;\anonymous \;\Varid{·}\;\anonymous )\;((),\anonymous ){}\<[E]%
\\
\>[3]{}\Varid{lookup}\;(\Conid{Pop}\;\Varid{i})\;(\anonymous \;\Varid{·}\;\Varid{env})\;(\anonymous ,\Varid{p})\;\mathrel{=}\;\Varid{lookup}\;\Varid{i}\;\Varid{env}\;\Varid{p}{}\<[E]%
\ColumnHook
\end{hscode}\resethooks
If the argument reference is \ensuremath{\Conid{Top}}, we pattern match on the
environment, which must contain a closure. We use the proof that the
environment contains exclusively closures to discharge the \ensuremath{\Conid{Clapp}}
branch. If the argument reference is \ensuremath{\Conid{Pop}\;\Varid{i}}, we recurse over \ensuremath{\Varid{i}} and
the tail of the environment.

Once again, we define a \ensuremath{\Conid{Trace}} data type, describing the call-graph
of the Krivine machine. The \ensuremath{\Conid{Trace}} data type is indexed by the three
arguments to the Krivine machine: a term, an environment, and an
evaluation context. The data type has a constructor for every
transition; recursive calls to the abstract machine correspond to
recursive arguments to a constructor:

\begin{hscode}\SaveRestoreHook
\column{B}{@{}>{\hspre}l<{\hspost}@{}}%
\column{3}{@{}>{\hspre}l<{\hspost}@{}}%
\column{5}{@{}>{\hspre}l<{\hspost}@{}}%
\column{7}{@{}>{\hspre}l<{\hspost}@{}}%
\column{E}{@{}>{\hspre}l<{\hspost}@{}}%
\>[3]{}\Keyword{data}\;\Conid{Trace}\;\mathbin{:}\;\hidden{\Gamma \;\sigma \;\tau }\;\Conid{Term}\;\Gamma \;\sigma \;\to \;\Conid{Env}\;\Gamma \;\to \;\Conid{EvalContext}\;\sigma \;\tau \;\to \;\Conid{Set}\;\Keyword{where}{}\<[E]%
\\
\>[3]{}\hsindent{2}{}\<[5]%
\>[5]{}\Conid{Lookup}\;\mathbin{:}\;\hidden{\sigma \;\tau \;\Gamma }\;\!\!\;\!\!\;(\Varid{i}\;\mathbin{:}\;\Conid{Ref}\;\Gamma \;\sigma )\;(\Varid{p}\;\mathbin{:}\;\Varid{isValidEnv}\;\Varid{env})\;\to {}\<[E]%
\\
\>[5]{}\hsindent{2}{}\<[7]%
\>[7]{}\Keyword{let}\;\Varid{c}\;\mathrel{=}\;\Varid{lookup}\;\Varid{i}\;\Varid{env}\;\Varid{p}\;\Keyword{in}{}\<[E]%
\\
\>[5]{}\hsindent{2}{}\<[7]%
\>[7]{}\Conid{Trace}\;(\Varid{getTerm}\;\Varid{c})\;(\Varid{getEnv}\;\Varid{c})\;\Varid{ctx}\;\to \;\Conid{Trace}\;(\Conid{Var}\;\Varid{i})\;\Varid{env}\;\Varid{ctx}{}\<[E]%
\\
\>[3]{}\hsindent{2}{}\<[5]%
\>[5]{}\Conid{App}\;\mathbin{:}\;\hidden{\sigma \;\Gamma \;\tau \;\rho }\;\!\!\;\!\!\;(\Varid{f}\;\mathbin{:}\;\Conid{Term}\;\Gamma \;(\sigma \;\Rightarrow\;\tau ))\;(\Varid{x}\;\mathbin{:}\;\Conid{Term}\;\Gamma \;\sigma )\;\to \;{}\<[E]%
\\
\>[5]{}\hsindent{2}{}\<[7]%
\>[7]{}\Conid{Trace}\;\Varid{f}\;\Varid{env}\;(\Conid{ARG}\;(\Conid{Closure}\;\Varid{x}\;\Varid{env})\;\Varid{ctx})\;\to \;{}\<[E]%
\\
\>[5]{}\hsindent{2}{}\<[7]%
\>[7]{}\Conid{Trace}\;(\Conid{App}\;\Varid{f}\;\Varid{x})\;\Varid{env}\;\Varid{ctx}{}\<[E]%
\\
\>[3]{}\hsindent{2}{}\<[5]%
\>[5]{}\Conid{Beta}\;\mathbin{:}\;\hidden{\Gamma \;\sigma \;\tau \;\rho \;\Conid{H}}\;\!\!\;(\Varid{ctx}\;\mathbin{:}\;\Conid{EvalContext}\;\sigma \;\rho )\;\to \;{}\<[E]%
\\
\>[5]{}\hsindent{2}{}\<[7]%
\>[7]{}(\Varid{arg}\;\mathbin{:}\;\Conid{Term}\;\Conid{H}\;\tau )\;\to \;(\Varid{argEnv}\;\mathbin{:}\;\Conid{Env}\;\Conid{H})\;\to \;{}\<[E]%
\\
\>[5]{}\hsindent{2}{}\<[7]%
\>[7]{}(\Varid{body}\;\mathbin{:}\;\Conid{Term}\;(\Conid{Cons}\;\tau \;\Gamma )\;\sigma )\;\to \;{}\<[E]%
\\
\>[5]{}\hsindent{2}{}\<[7]%
\>[7]{}\Conid{Trace}\;\Varid{body}\;(\Conid{Closure}\;\Varid{arg}\;\Varid{argEnv}\;\Varid{·}\;\Varid{env})\;\Varid{ctx}\;\to \;{}\<[E]%
\\
\>[5]{}\hsindent{2}{}\<[7]%
\>[7]{}\Conid{Trace}\;(\Conid{Lam}\;\Varid{body})\;\Varid{env}\;(\Conid{ARG}\;(\Conid{Closure}\;\Varid{arg}\;\Varid{argEnv})\;\Varid{ctx}){}\<[E]%
\\
\>[3]{}\hsindent{2}{}\<[5]%
\>[5]{}\Conid{Done}\;\mathbin{:}\;\hidden{\Gamma \;\sigma \;\tau }\;\!\!\;(\Varid{body}\;\mathbin{:}\;\Conid{Term}\;(\Conid{Cons}\;\tau \;\Gamma )\;\sigma )\;\to \;\Conid{Trace}\;(\Conid{Lam}\;\Varid{body})\;\Varid{env}\;\Conid{MT}{}\<[E]%
\ColumnHook
\end{hscode}\resethooks

Using this \ensuremath{\Conid{Trace}}, we can now define the final version of the
\ensuremath{\Varid{refocus}} function, corresponding to the Krivine abstract machine, by
structural recursion on this \ensuremath{\Conid{Trace}}. The resulting machine
corresponds to the Krivine machine as is usually presented in the
literature~\cite{cregut,curien,hankin}. Biernacka and
Danvy~\cite{biernacka} also consider the derivation of Krivine's
original machine~\cite{krivine} that contracts nested
$\beta$-reductions in one step.

\begin{hscode}\SaveRestoreHook
\column{B}{@{}>{\hspre}l<{\hspost}@{}}%
\column{3}{@{}>{\hspre}l<{\hspost}@{}}%
\column{4}{@{}>{\hspre}l<{\hspost}@{}}%
\column{5}{@{}>{\hspre}l<{\hspost}@{}}%
\column{E}{@{}>{\hspre}l<{\hspost}@{}}%
\>[3]{}\Varid{refocus}\;\mathbin{:}\;\hidden{\Gamma \;\sigma \;\tau }\;(\Varid{ctx}\;\mathbin{:}\;\Conid{EvalContext}\;\sigma \;\tau )\;(\Varid{t}\;\mathbin{:}\;\Conid{Term}\;\Gamma \;\sigma )\;(\Varid{env}\;\mathbin{:}\;\Conid{Env}\;\Gamma )\;\to \;{}\<[E]%
\\
\>[3]{}\hsindent{1}{}\<[4]%
\>[4]{}\Conid{Trace}\;\Varid{t}\;\Varid{env}\;\Varid{ctx}\;\to \;\Conid{Value}\;\tau {}\<[E]%
\\
\>[3]{}\Varid{refocus}\;\Varid{ctx}\;\lfloor \Conid{Var}\;\Varid{i}\rfloor\;\Varid{env}\;(\Conid{Lookup}\;\Varid{i}\;\Varid{q}\;\Varid{step})\;\mathrel{=}{}\<[E]%
\\
\>[3]{}\hsindent{2}{}\<[5]%
\>[5]{}\Keyword{let}\;\Varid{c}\;\mathrel{=}\;\Varid{lookup}\;\Varid{i}\;\Varid{env}\;\Varid{q}\;\Keyword{in}{}\<[E]%
\\
\>[3]{}\hsindent{2}{}\<[5]%
\>[5]{}\Varid{refocus}\;\Varid{ctx}\;(\Varid{getTerm}\;\Varid{c})\;(\Varid{getEnv}\;\Varid{c})\;\Varid{step}{}\<[E]%
\\
\>[3]{}\Varid{refocus}\;\Varid{ctx}\;\lfloor \Conid{App}\;\Varid{f}\;\Varid{x}\rfloor\;\Varid{env}\;(\Conid{App}\;\Varid{f}\;\Varid{x}\;\Varid{step})\;{}\<[E]%
\\
\>[3]{}\hsindent{2}{}\<[5]%
\>[5]{}\mathrel{=}\;\Varid{refocus}\;(\Conid{ARG}\;(\Conid{Closure}\;\Varid{x}\;\Varid{env})\;\Varid{ctx})\;\Varid{f}\;\Varid{env}\;\Varid{step}{}\<[E]%
\\
\>[3]{}\Varid{refocus}\;\lfloor \Conid{ARG}\;(\Conid{Closure}\;\Varid{arg}\;\Varid{env'})\;\Varid{ctx}\rfloor\;\lfloor \Conid{Lam}\;\Varid{body}\rfloor\;\Varid{env}\;(\Conid{Beta}\;\Varid{ctx}\;\Varid{arg}\;\Varid{env'}\;\Varid{body}\;\Varid{step})\;{}\<[E]%
\\
\>[3]{}\hsindent{2}{}\<[5]%
\>[5]{}\mathrel{=}\;\Varid{refocus}\;\Varid{ctx}\;\Varid{body}\;((\Conid{Closure}\;\Varid{arg}\;\Varid{env'})\;\Varid{·}\;\Varid{env})\;\Varid{step}{}\<[E]%
\\
\>[3]{}\Varid{refocus}\;\lfloor \Conid{MT}\rfloor\;\lfloor \Conid{Lam}\;\Varid{body}\rfloor\;\Varid{env}\;(\Conid{Done}\;\Varid{body})\;\mathrel{=}\;\Conid{Val}\;(\Conid{Closure}\;(\Conid{Lam}\;\Varid{body})\;\Varid{env})\;\Varid{unit}{}\<[E]%
\ColumnHook
\end{hscode}\resethooks
In the case for variables, we look up the closure that the variable
refers to in the environment, and continue evaluation with that closure's
term and environment. In the case for \ensuremath{\Conid{App}\;\Varid{f}\;\Varid{x}}, we add the argument
and current environment to the application context, and continue
evaluating the term \ensuremath{\Varid{f}}. We distinguish two further cases for lambda
terms: if the evaluation context is not empty, we can perform a beta
reduction step; otherwise evaluation is finished.

We still need to prove that the \ensuremath{\Conid{Trace}} data type is inhabited. During
execution, the Krivine machine only adds closures to the environment
and evaluation context. During the termination proof, we will need to
keep track of the following invariant on evaluation contexts and
environments:
\begin{hscode}\SaveRestoreHook
\column{B}{@{}>{\hspre}l<{\hspost}@{}}%
\column{3}{@{}>{\hspre}l<{\hspost}@{}}%
\column{E}{@{}>{\hspre}l<{\hspost}@{}}%
\>[3]{}\Varid{invariant}\;\mathbin{:}\;\hidden{\Gamma \;\sigma \;\tau }\;\Conid{EvalContext}\;\sigma \;\tau \;\to \;\Conid{Env}\;\Gamma \;\to \;\Conid{Set}{}\<[E]%
\\
\>[3]{}\Varid{invariant}\;\Varid{ctx}\;\Varid{env}\;\mathrel{=}\;\Conid{Pair}\;(\Varid{isValidEnv}\;\Varid{env})\;(\Varid{isValidContext}\;\Varid{ctx}){}\<[E]%
\ColumnHook
\end{hscode}\resethooks
The proof of termination once again calls the \ensuremath{\Varid{termination}} proof from
the previous section. An auxiliary lemma shows that any witness of
termination for the small-step abstract machine in
Section~\ref{sec:refocusing} will also suffice as a proof of
termination of the Krivine machine.

\begin{hscode}\SaveRestoreHook
\column{B}{@{}>{\hspre}l<{\hspost}@{}}%
\column{3}{@{}>{\hspre}l<{\hspost}@{}}%
\column{5}{@{}>{\hspre}l<{\hspost}@{}}%
\column{7}{@{}>{\hspre}l<{\hspost}@{}}%
\column{E}{@{}>{\hspre}l<{\hspost}@{}}%
\>[3]{}\Varid{termination}\;\mathbin{:}\;\hidden{\sigma }\;(\Varid{t}\;\mathbin{:}\;\Conid{Term}\;\Conid{Nil}\;\sigma )\;\to \;\Conid{Trace}\;\Varid{t}\;\Conid{Nil}\;\Conid{MT}{}\<[E]%
\\
\>[3]{}\Varid{termination}\;\Varid{t}\;\mathrel{=}\;\Varid{lemma}\;\Conid{MT}\;\Varid{t}\;\Conid{Nil}\;(\Varid{unit},\Varid{unit})\;(\Conid{Section5.termination}\;(\Conid{Closure}\;\Varid{t}\;\Conid{Nil})){}\<[E]%
\\
\>[3]{}\hsindent{2}{}\<[5]%
\>[5]{}\Keyword{where}{}\<[E]%
\\
\>[3]{}\hsindent{2}{}\<[5]%
\>[5]{}\Varid{lemma}\;\mathbin{:}\;\hidden{\Gamma \;\sigma \;\tau }\;(\Varid{ctx}\;\mathbin{:}\;\Conid{EvalContext}\;\sigma \;\tau )\;(\Varid{t}\;\mathbin{:}\;\Conid{Term}\;\Gamma \;\sigma )\;(\Varid{env}\;\mathbin{:}\;\Conid{Env}\;\Gamma )\;\to \;{}\<[E]%
\\
\>[5]{}\hsindent{2}{}\<[7]%
\>[7]{}\Varid{invariant}\;\Varid{ctx}\;\Varid{env}\;\to \;\Conid{Section5.Trace}\;(\Conid{Section5.refocus}\;\Varid{ctx}\;(\Conid{Closure}\;\Varid{t}\;\Varid{env}))\;\to \;{}\<[E]%
\\
\>[5]{}\hsindent{2}{}\<[7]%
\>[7]{}\Conid{Trace}\;\Varid{t}\;\Varid{env}\;\Varid{ctx}{}\<[E]%
\ColumnHook
\end{hscode}\resethooks
The lemma is proven by straightforward induction on the evaluation
context, the term, and the \ensuremath{\Conid{Trace}} data type from the previous
section. Once we pattern match on the term and the evaluation context,
we know which transition we wish to make, and hence which constructor
of the \ensuremath{\Conid{Trace}} data type is required. Any recursive occurrences of the
\ensuremath{\Conid{Trace}} data type can be produced by recursive calls to the
\ensuremath{\Varid{lemma}}. The only other result necessary states that the \ensuremath{\Varid{lookup}}
function and the \ensuremath{\Varid{\char95 !\char95 }} operation we saw previously return the same
closed term from an environment.

Finally, we can define the \ensuremath{\Varid{evaluation}} function that calls \ensuremath{\Varid{refocus}}
with a suitable choice for its initial arguments:
\begin{hscode}\SaveRestoreHook
\column{B}{@{}>{\hspre}l<{\hspost}@{}}%
\column{3}{@{}>{\hspre}l<{\hspost}@{}}%
\column{E}{@{}>{\hspre}l<{\hspost}@{}}%
\>[3]{}\Varid{evaluate}\;\mathbin{:}\;\hidden{\sigma }\;\Conid{Term}\;\Conid{Nil}\;\sigma \;\to \;\Conid{Value}\;\sigma {}\<[E]%
\\
\>[3]{}\Varid{evaluate}\;\Varid{t}\;\mathrel{=}\;\Varid{refocus}\;\Conid{MT}\;\Varid{t}\;\Conid{Nil}\;(\Varid{termination}\;\Varid{t}){}\<[E]%
\ColumnHook
\end{hscode}\resethooks

To conclude, we show that this final version of the \ensuremath{\Varid{refocus}} function
behaves equivalently to the \ensuremath{\Varid{refocus}} function from the previous
section. To prove this, we formulate the correctness property below.

\begin{hscode}\SaveRestoreHook
\column{B}{@{}>{\hspre}l<{\hspost}@{}}%
\column{3}{@{}>{\hspre}l<{\hspost}@{}}%
\column{4}{@{}>{\hspre}l<{\hspost}@{}}%
\column{18}{@{}>{\hspre}l<{\hspost}@{}}%
\column{E}{@{}>{\hspre}l<{\hspost}@{}}%
\>[3]{}\Varid{correctness}\;\mathbin{:}\;{}\<[18]%
\>[18]{}\hidden{\sigma \;\tau \;\Gamma }\;(\Varid{ctx}\;\mathbin{:}\;\Conid{EvalContext}\;\sigma \;\tau )\;(\Varid{t}\;\mathbin{:}\;\Conid{Term}\;\Gamma \;\sigma )\;(\Varid{env}\;\mathbin{:}\;\Conid{Env}\;\Gamma )\;\to \;{}\<[E]%
\\
\>[3]{}\hsindent{1}{}\<[4]%
\>[4]{}(\Varid{t}_{\Varid{1}}\;\mathbin{:}\;\Conid{Trace}\;\Varid{t}\;\Varid{env}\;\Varid{ctx})\;\to \;{}\<[E]%
\\
\>[3]{}\hsindent{1}{}\<[4]%
\>[4]{}(\Varid{t}_{\Varid{2}}\;\mathbin{:}\;\Conid{Section5.Trace}\;(\Conid{Section5.refocus}\;\Varid{ctx}\;(\Conid{Closure}\;\Varid{t}\;\Varid{env})))\;\to \;{}\<[E]%
\\
\>[3]{}\hsindent{1}{}\<[4]%
\>[4]{}\Varid{refocus}\;\Varid{ctx}\;\Varid{t}\;\Varid{env}\;\Varid{t}_{\Varid{1}}\;\equiv \;\Conid{Section5.iterate}\;(\Conid{Section5.refocus}\;\Varid{ctx}\;(\Conid{Closure}\;\Varid{t}\;\Varid{env}))\;\Varid{t}_{\Varid{2}}{}\<[E]%
\ColumnHook
\end{hscode}\resethooks
Once again, the proof proceeds by straightforward induction on the traces.

As a result of this correctness property, we can prove that our
evaluation function behaves the same as the function presented in the
previous section:
\begin{hscode}\SaveRestoreHook
\column{B}{@{}>{\hspre}l<{\hspost}@{}}%
\column{3}{@{}>{\hspre}l<{\hspost}@{}}%
\column{18}{@{}>{\hspre}l<{\hspost}@{}}%
\column{30}{@{}>{\hspre}l<{\hspost}@{}}%
\column{E}{@{}>{\hspre}l<{\hspost}@{}}%
\>[3]{}\Varid{corollary}\;\mathbin{:}\;\hidden{\sigma }\;(\Varid{t}\;\mathbin{:}\;\Conid{Term}\;\Conid{Nil}\;\sigma )\;\to \;\Varid{evaluate}\;\Varid{t}\;\equiv \;\Conid{Section5.evaluate}\;(\Conid{Closure}\;\Varid{t}\;\Conid{Nil}){}\<[E]%
\\
\>[3]{}\Varid{corollary}\;\Varid{t}\;\mathrel{=}\;{}\<[18]%
\>[18]{}\Keyword{let}\;\Varid{trace}\;{}\<[30]%
\>[30]{}\mathrel{=}\;\Varid{termination}\;\Varid{t}\;\Keyword{in}{}\<[E]%
\\
\>[18]{}\Keyword{let}\;\Varid{trace'}\;{}\<[30]%
\>[30]{}\mathrel{=}\;\Conid{Section5.termination}\;(\Conid{Closure}\;\Varid{t}\;\Conid{Nil})\;\Keyword{in}{}\<[E]%
\\
\>[18]{}\Varid{correctness}\;\Conid{MT}\;\Varid{t}\;\Conid{Nil}\;\Varid{trace}\;\Varid{trace'}{}\<[E]%
\ColumnHook
\end{hscode}\resethooks

By chaining together our correctness results, we can show that our
Krivine machine produces the same value as our original evaluator
based on repeated head reduction, thereby completing the formal
derivation of the Krivine machine from a small step evaluator.

\section{Discussion}
\label{sec:discussion}

There has been previous work on formalizing the derivations of
abstract machines in Coq~\cite{biernackis,filip}. In contrast to the
development here, these formalizations are not executable but instead
define the reduction behaviour as inductive relations between terms
and values. The executability of our abstract machines comes at a
price: we need to prove that the evaluators terminate, which requires
a clever logical relation. On the other hand, it is easier to reason
about executable functions. In type theory, definitional equalities
are always trivially true---a fact you can only exploit if your
functions compute.

This paper uses the Bove-Capretta method to prove termination of every
evaluator. Chapman and Altenkirch use a similar logical relation to
produce inhabitants of Bove-Capretta predicates when writing a
big-step normalization algorithm~\cite{chapmeister}. There are, of
course, alternative methods to show that a non-structurally recursive
function does terminate. For example, it may be interesting to
investigate how to adapt the normalization proof to use an order on
lambda terms proposed by Gandy~\cite{gandy} to define a suitable
accessibility relation.

Finally, you may wonder if the usage of logical relations to prove
termination is `cheating.' After all, the computational content of
normalization proofs using logical relations is itself a normalization
algorithm~\cite{berger,extraction,stovring}---so is our small-step
evaluator not just reading off the value from the trace that our proof
computes? Not at all! In fact, the behaviour of the \ensuremath{\Varid{iterate}} function
from Section~\ref{sec:iterated-head-reduction} is \emph{independent}
of the trace we provide---once the \ensuremath{\Varid{iterate}} function matches on the
argument decomposition, the trace passed as an argument to the
\ensuremath{\Varid{iterate}} function is uniquely determined. The following statement is
easy to prove:
\begin{hscode}\SaveRestoreHook
\column{B}{@{}>{\hspre}l<{\hspost}@{}}%
\column{5}{@{}>{\hspre}l<{\hspost}@{}}%
\column{E}{@{}>{\hspre}l<{\hspost}@{}}%
\>[5]{}\Varid{collapsible}\;\mathbin{:}\;(\Varid{d}\;\mathbin{:}\;\Conid{Decomposition}\;\Varid{c})\;(\Varid{t}_{\Varid{1}}\;\Varid{t}_{\Varid{2}}\;\mathbin{:}\;\Conid{Trace}\;\Varid{d})\;\to \;\Varid{t}_{\Varid{1}}\;\equiv \;\Varid{t}_{\Varid{2}}{}\<[E]%
\ColumnHook
\end{hscode}\resethooks
In other words, the traces themselves carry no computational
content. Such \emph{collapsible} data types may be erased by a
suitable clever compiler~\cite{edwinI,edwinII}.

This paper focuses on the derivation of the Krivine abstract
machine. There is no reason to believe that the other derivations of
abstract machines~\cite{ager,biernacka} may not be formalized in a
similar fashion.

\subsection*{Acknowledgements}
\label{sec:acks}

I would like to thank James McKinna for our entertaining and
educational discussions. Małgorzata Biernacka, Pierre-Evariste Dagand,
Olivier Danvy, Ilya Sergey, Thomas van Noort and four anonymous
reviewers all provided invaluable feedback on a draft version of this
paper, for which I am grateful.

\bibliographystyle{eptcs}
\bibliography{M2M}

\appendix

\section{An Agda Prelude}
\label{appendix}

\begin{hscode}\SaveRestoreHook
\column{B}{@{}>{\hspre}l<{\hspost}@{}}%
\column{3}{@{}>{\hspre}l<{\hspost}@{}}%
\column{5}{@{}>{\hspre}l<{\hspost}@{}}%
\column{E}{@{}>{\hspre}l<{\hspost}@{}}%
\>[B]{}\Keyword{module}\;\Conid{Prelude}\;\Keyword{where}{}\<[E]%
\\[\blanklineskip]%
\>[B]{}\hsindent{3}{}\<[3]%
\>[3]{}\Varid{id}\;\mathbin{:}\;\Keyword{forall}\;\{\mskip1.5mu \Varid{a}\;\mathbin{:}\;\Conid{Set}\mskip1.5mu\}\;\to \;\Varid{a}\;\to \;\Varid{a}{}\<[E]%
\\
\>[B]{}\hsindent{3}{}\<[3]%
\>[3]{}\Varid{id}\;\Varid{x}\;\mathrel{=}\;\Varid{x}{}\<[E]%
\\[\blanklineskip]%
\>[B]{}\hsindent{3}{}\<[3]%
\>[3]{}\Keyword{data}\;\Conid{Empty}\;\mathbin{:}\;\Conid{Set}\;\Keyword{where}{}\<[E]%
\\[\blanklineskip]%
\>[B]{}\hsindent{3}{}\<[3]%
\>[3]{}\Varid{magic}\;\mathbin{:}\;\Keyword{forall}\;\{\mskip1.5mu \Varid{a}\;\mathbin{:}\;\Conid{Set}\mskip1.5mu\}\;\to \;\Conid{Empty}\;\to \;\Varid{a}{}\<[E]%
\\
\>[B]{}\hsindent{3}{}\<[3]%
\>[3]{}\Varid{magic}\;(){}\<[E]%
\\[\blanklineskip]%
\>[B]{}\hsindent{3}{}\<[3]%
\>[3]{}\Keyword{record}\;\Conid{Unit}\;\mathbin{:}\;\Conid{Set}\;\Keyword{where}{}\<[E]%
\\[\blanklineskip]%
\>[B]{}\hsindent{3}{}\<[3]%
\>[3]{}\Varid{unit}\;\mathbin{:}\;\Conid{Unit}{}\<[E]%
\\
\>[B]{}\hsindent{3}{}\<[3]%
\>[3]{}\Varid{unit}\;\mathrel{=}\;\Keyword{record}\;\{\mskip1.5mu \mskip1.5mu\}{}\<[E]%
\\[\blanklineskip]%
\>[B]{}\hsindent{3}{}\<[3]%
\>[3]{}\Keyword{data}\;\Conid{Pair}\;(\Varid{a}\;\Varid{b}\;\mathbin{:}\;\Conid{Set})\;\mathbin{:}\;\Conid{Set}\;\Keyword{where}{}\<[E]%
\\
\>[3]{}\hsindent{2}{}\<[5]%
\>[5]{}\anonymous ,\anonymous \;\mathbin{:}\;\Varid{a}\;\to \;\Varid{b}\;\to \;\Conid{Pair}\;\Varid{a}\;\Varid{b}{}\<[E]%
\\[\blanklineskip]%
\>[B]{}\hsindent{3}{}\<[3]%
\>[3]{}\Varid{fst}\;\mathbin{:}\;\Keyword{forall}\;\{\mskip1.5mu \Varid{a}\;\Varid{b}\mskip1.5mu\}\;\to \;\Conid{Pair}\;\Varid{a}\;\Varid{b}\;\to \;\Varid{a}{}\<[E]%
\\
\>[B]{}\hsindent{3}{}\<[3]%
\>[3]{}\Varid{fst}\;(\Varid{x},\anonymous )\;\mathrel{=}\;\Varid{x}{}\<[E]%
\\[\blanklineskip]%
\>[B]{}\hsindent{3}{}\<[3]%
\>[3]{}\Varid{snd}\;\mathbin{:}\;\Keyword{forall}\;\{\mskip1.5mu \Varid{a}\;\Varid{b}\mskip1.5mu\}\;\to \;\Conid{Pair}\;\Varid{a}\;\Varid{b}\;\to \;\Varid{b}{}\<[E]%
\\
\>[B]{}\hsindent{3}{}\<[3]%
\>[3]{}\Varid{snd}\;(\anonymous ,\Varid{y})\;\mathrel{=}\;\Varid{y}{}\<[E]%
\\[\blanklineskip]%
\>[B]{}\hsindent{3}{}\<[3]%
\>[3]{}\Keyword{data}\;\Conid{List}\;(\Varid{a}\;\mathbin{:}\;\Conid{Set})\;\mathbin{:}\;\Conid{Set}\;\Keyword{where}{}\<[E]%
\\
\>[3]{}\hsindent{2}{}\<[5]%
\>[5]{}\Conid{Nil}\;\mathbin{:}\;\Conid{List}\;\Varid{a}{}\<[E]%
\\
\>[3]{}\hsindent{2}{}\<[5]%
\>[5]{}\Conid{Cons}\;\mathbin{:}\;\Varid{a}\;\to \;\Conid{List}\;\Varid{a}\;\to \;\Conid{List}\;\Varid{a}{}\<[E]%
\\[\blanklineskip]%
\>[B]{}\hsindent{3}{}\<[3]%
\>[3]{}\Keyword{data}\;\_\equiv\_\;\{\mskip1.5mu \Varid{a}\;\mathbin{:}\;\Conid{Set}\mskip1.5mu\}\;(\Varid{x}\;\mathbin{:}\;\Varid{a})\;\mathbin{:}\;\Varid{a}\;\to \;\Conid{Set}\;\Keyword{where}{}\<[E]%
\\
\>[3]{}\hsindent{2}{}\<[5]%
\>[5]{}\Conid{Refl}\;\mathbin{:}\;\Varid{x}\;\equiv \;\Varid{x}{}\<[E]%
\\[\blanklineskip]%
\>[B]{}\hsindent{3}{}\<[3]%
\>[3]{}\Keyword{infix}\;\Varid{6}\;\_\equiv\_{}\<[E]%
\\[\blanklineskip]%
\>[B]{}\hsindent{3}{}\<[3]%
\>[3]{}\Varid{sym}\;\mathbin{:}\;\{\mskip1.5mu \Varid{a}\;\mathbin{:}\;\Conid{Set}\mskip1.5mu\}\;\{\mskip1.5mu \Varid{x}\;\Varid{y}\;\mathbin{:}\;\Varid{a}\mskip1.5mu\}\;\to \;\Varid{x}\;\equiv \;\Varid{y}\;\to \;\Varid{y}\;\equiv \;\Varid{x}{}\<[E]%
\\
\>[B]{}\hsindent{3}{}\<[3]%
\>[3]{}\Varid{sym}\;\Conid{Refl}\;\mathrel{=}\;\Conid{Refl}{}\<[E]%
\\[\blanklineskip]%
\>[B]{}\hsindent{3}{}\<[3]%
\>[3]{}\Varid{cong}\;\mathbin{:}\;\{\mskip1.5mu \Varid{a}\;\Varid{b}\;\mathbin{:}\;\Conid{Set}\mskip1.5mu\}\;\{\mskip1.5mu \Varid{x}\;\Varid{y}\;\mathbin{:}\;\Varid{a}\mskip1.5mu\}\;\to \;(\Varid{f}\;\mathbin{:}\;\Varid{a}\;\to \;\Varid{b})\;\to \;\Varid{x}\;\equiv \;\Varid{y}\;\to \;\Varid{f}\;\Varid{x}\;\equiv \;\Varid{f}\;\Varid{y}{}\<[E]%
\\
\>[B]{}\hsindent{3}{}\<[3]%
\>[3]{}\Varid{cong}\;\Varid{f}\;\Conid{Refl}\;\mathrel{=}\;\Conid{Refl}{}\<[E]%
\\[\blanklineskip]%
\>[B]{}\hsindent{3}{}\<[3]%
\>[3]{}\Keyword{data}\;\Conid{Exists}\;(\Varid{a}\;\mathbin{:}\;\Conid{Set})\;(\Varid{b}\;\mathbin{:}\;\Varid{a}\;\to \;\Conid{Set})\;\mathbin{:}\;\Conid{Set}\;\Keyword{where}{}\<[E]%
\\
\>[3]{}\hsindent{2}{}\<[5]%
\>[5]{}\Conid{Witness}\;\mathbin{:}\;(\Varid{x}\;\mathbin{:}\;\Varid{a})\;\to \;\Varid{b}\;\Varid{x}\;\to \;\Conid{Exists}\;\Varid{a}\;\Varid{b}{}\<[E]%
\\[\blanklineskip]%
\>[B]{}\hsindent{3}{}\<[3]%
\>[3]{}\Varid{fsts}\;\mathbin{:}\;\Keyword{forall}\;\{\mskip1.5mu \Varid{a}\;\Varid{b}\mskip1.5mu\}\;\to \;\Conid{Exists}\;\Varid{a}\;\Varid{b}\;\to \;\Varid{a}{}\<[E]%
\\
\>[B]{}\hsindent{3}{}\<[3]%
\>[3]{}\Varid{fsts}\;(\Conid{Witness}\;\Varid{x}\;\anonymous )\;\mathrel{=}\;\Varid{x}{}\<[E]%
\\[\blanklineskip]%
\>[B]{}\hsindent{3}{}\<[3]%
\>[3]{}\Varid{snds}\;\mathbin{:}\;\Keyword{forall}\;\{\mskip1.5mu \Varid{a}\;\Varid{b}\mskip1.5mu\}\;\to \;(\Varid{x}\;\mathbin{:}\;\Conid{Exists}\;\Varid{a}\;\Varid{b})\;\to \;(\Varid{b}\;(\Varid{fsts}\;\Varid{x})){}\<[E]%
\\
\>[B]{}\hsindent{3}{}\<[3]%
\>[3]{}\Varid{snds}\;(\Conid{Witness}\;\anonymous \;\Varid{y})\;\mathrel{=}\;\Varid{y}{}\<[E]%
\ColumnHook
\end{hscode}\resethooks

\end{document}